\documentclass[]{aa}  
\usepackage{array, multirow, graphicx}
\usepackage{wasysym}
\usepackage{siunitx}
\usepackage{booktabs, tabularx}

\usepackage{txfonts}
\usepackage[]{hyperref}
\hypersetup{colorlinks,citecolor=blue, linkcolor=blue, urlcolor=blue, breaklinks=True}
\usepackage{breakcites}
\newcounter{magicrownumbers}

\usepackage{cellspace, makecell} 
\usepackage{mathtools}

\begin{document} 
\newcommand{\bcdot}{\boldsymbol{\cdot}}
\newcommand{\w}[1]{\mathbf{#1}}
\newcommand{\bra}[1]{\left\langle #1 \middle | \right.}
\newcommand{\ket}[1]{\left.\middle | #1\right\rangle}
\newcommand{\braket}[2]{\left\langle #1  \middle | #2 \right\rangle}
\newcommand{\oplin}[2]{\left.\middle | #1 \right \rangle \left \langle #2 \middle | \right.}
\newcommand{\lf}{\left}
\newcommand{\rg}{\right}
\newcommand{\tonda}[1]{\!\left ( #1 \right )}
\newcommand{\quadra}[1]{\left [ #1 \right ]}
\newcommand{\graffa}[1]{\left \{ #1 \right \}}
\newcommand{\dirac}[1]{\updelta\!\left( #1 \right )}
\newcommand{\deltak}[1]{\updelta_{ #1 }}
\newcommand{\D}[2]{\frac{d #1}{d  #2}}
\newcommand{\DD}[2]{\frac{d^2 #1}{d #2^2}}
\newcommand{\pd}[2]{\frac{\partial #1}{\partial #2}}
\newcommand{\pdd}[2]{\frac{\partial^2 #1}{\partial #2^2}}
\newcommand{\medio}[1]{\left\langle #1 \right\rangle}
\newcommand{\abs}[1]{\left | #1 \right |}
\newcommand{\norma}[1]{\left \| #1 \right \|}
\newcommand{\sca}[2]{\left \langle #1 , #2 \right\rangle}
\newcommand{\ndiv}[1]{{\boldsymbol{\nabla}}\boldsymbol{\cdot}\w{#1}}
\newcommand{\ndivx}[1]{{\boldsymbol{\nabla}_{\w{x}}}\boldsymbol{\cdot}\w{#1}}
\newcommand{\ndivv}[1]{{\boldsymbol{\nabla}_{\w{v}}}\boldsymbol{\cdot}\w{#1}}
\newcommand{\grad}{{\boldsymbol{\nabla}}}
\newcommand{\gradx}{{\boldsymbol{\nabla}}_{\w{x}}}
\newcommand{\gradv}{{\boldsymbol{\nabla}}_{\w{v}}}
\newcommand{\lap}[1]{\boldsymbol{\nabla}^2 #1}
\newcommand{\rot}[1]{\boldsymbol{\nabla}\times\w{#1}}
\newcommand{\pois}[2]{\Bigl \{ #1 , #2 \Bigr\}}
\newcommand{\com}[2]{\left [ #1 , #2 \right ]}
\newcommand{\sistemai}{\lf\{\begin{aligned}}
\newcommand{\sistemaf}{\end{aligned}\rg.}
\newcommand{\implica}{\Longrightarrow}
\newcommand{\sse}{\Longleftrightarrow}
\newcommand{\id}{\mathbb{I}}
\newcommand{\sopra}[1]{\overline{#1}}
\newcommand{\sotto}[1]{\underline{#1}}
\newcommand{\gv}[1]{\ensuremath{\mbox{\boldmath$ #1 $}}}
\newcommand{\calc}[3]{\lf.{#1}\rg |_{#2}^{#3}}
\newcommand{\cro}{\dagger}
\newcommand{\eq}[1]{\begin{equation} #1 \end{equation}}
\newcommand{\eqn}[1]{\begin{equation*} #1 \end{equation*}}
\newcommand{\virg}[1]{``#1''}
\newcommand{\dmat}[1]{\frac{\mathcal{D} #1}{\mathcal{D}t}}
\newcommand{\mc}[1]{\mathcal{#1}}
\newcommand{\bxi}{\boldsymbol{\xi}}
\newcommand{\vx}{\hat{\w{x}}}
\newcommand{\vy}{\hat{\w{y}}}
\newcommand{\vz}{\hat{\w{z}}}
\newcommand{\vn}{\hat{\w{n}}}
\newcommand{\disc}[1]{\biggl[ #1 \biggr]}
\newcommand{\td}{\tau_{\text{diff}}}
\newcommand{\tc}{\tau_{\text{conv}}}
\newcommand{\ft}{\tilde{f}_1}
\newcommand{\fth}{\widehat{f}_1}
\newcommand{\et}{\tilde{\w{E}}_1}
\newcommand{\eet}{\tilde{E}_1}
\newcommand{\eeth}{\widehat{E}_1}
\newcommand{\asun}{\astrosun}
\newcommand{\al}[1]{\begin{aligned} #1 \end{aligned}}
\newcommand{\ftonda}[2]{\!\left ( \frac{#1}{#2} \right )}
\newcommand{\mum}{\unit{\mu m}}
\newcommand{\hh}{H$_{2}$ }
\newcommand{\Mbh}{M_\text{BH}}
\newcommand{\bs}[1]{\boldsymbol{#1}}
\newcommand{\fit}{\emph{fit}\xspace}
\newcommand{\pixel}{\emph{pixel}\xspace}
\newcommand{\RNum}[1]{\uppercase\expandafter{\romannumeral #1\relax}}
\newcommand{\oi}{[O\RNum{1}]$_{\rm 63\,\mu m}$}
\newcommand{\oiii}[1]{[O\RNum{3}]$_{\rm #1\,\mu m}$}
\newcommand{\nii}[1]{[N\RNum{2}]$_{\rm #1\,\mu m}$}
\newcommand{\cii}{[C\RNum{2}]$_{\rm 158\,\mu m}$}
\newcommand{\ci}[1]{[C\RNum{1}]$_{\rm #1\,\mu m}$}

\newcommand{\red}[1]{\textcolor{red}{#1}}

   \title{Unveiling the warm dense ISM in $z>6$ quasar host galaxies\\ via water vapor emission}


   \author{A.~Pensabene
          \inst{\ref{unimib},\ref{difa},\ref{inaf-bo}}
         	\and
	P. van der Werf
	\inst{\ref{leiden}}
         	\and
         R.~Decarli
          \inst{\ref{inaf-bo}}
          \and
          E.~Ba{\~{n}}ados
          \inst{\ref{mpia}}
          \and
          R.~A.~Meyer
          \inst{\ref{mpia}}
          \and
	 D.~Riechers
	 \inst{\ref{uni-koln}}
	 \and
          B.~Venemans
          \inst{\ref{leiden}}
          \and
	  F.~Walter
	  \inst{\ref{mpia}, \ref{nrao}}
	  \and
	  A.~Wei{\ss}
	 \inst{\ref{mpi-bonn}}
	  \and
          M.~Brusa
          \inst{\ref{difa}}
          \and
          X.~Fan
          \inst{\ref{uni-az}}
          \and
          F.~Wang
          \inst{\ref{uni-az}}
          \and
          J.~Yang
          \inst{\ref{uni-az}}
          }

   \institute{Dipartimento di Fisica ``G. Occhialini'', Universit\`a degli Studi di Milano-Bicocca, Piazza della Scienza 3, I-20126, Milano, Italy\\\email{antonio.pensabene@unimib.it}\label{unimib}
   		\and
   Dipartimento di Fisica e Astronomia, Alma Mater Studiorum, Universit\`a di Bologna, Via Gobetti 93/2, I-40129 Bologna, Italy\label{difa}
   		\and
   	INAF-Osservatorio di Astrofisica e Scienza dello Spazio, Via Gobetti 93/3, I-40129 Bologna, Italy\label{inaf-bo}
		\and
		Leiden Observatory, Leiden University, P.O. box 9513, NL-2300 RA Leiden, the Netherlands\label{leiden}
		\and
	Max-Planck-Institut f\"{u}r Astronomie, K\"{o}nigstuhl 17, D-69117 Heidelberg, Germany\label{mpia}
		\and
	I. Physikalisches Institut, Universit\"at zu K\"oln, Z\"ulpicher Strasse 77, D-50937 K\"oln, Germany\label{uni-koln}
		\and
	National Radio Astronomy Observatory, Pete V. Domenici Array Science Center, P.O. Box O, Socorro, NM 87801, USA\label{nrao}
		\and
	Max-Planck-Institut f\"ur Radioastronomie, Auf dem H\"ugel 69, D-53121 Bonn, Germany\label{mpi-bonn}
		\and
	Steward Observatory, University of Arizona, 933 North Cherry Avenue, Tucson, AZ 85721, USA\label{uni-az}
             }

   \date{Received XXX accepted YYY}

 
\abstract{Water vapor (H$_{2}$O) is one of the brightest molecular emitters after carbon monoxide (CO) in galaxies with high infrared (IR) luminosity, and allows us to investigate the warm dense phase of the interstellar medium (ISM) where star formation occurs. However, due to the complexity of its radiative spectrum, H$_{2}$O is not frequently exploited as an ISM tracer in distant galaxies. Therefore, H$_{2}$O studies of the warm and dense gas at high-$z$ remains largely unexplored. %
In this work we present observations conducted with the Northern Extended Millimeter Array (NOEMA) toward three $z>6$ IR-bright quasars J2310+1855, J1148+5251, and J0439+1634 targeted in their multiple para-/ortho-H$_{2}$O transitions ($3_{12}-3_{03}$, $1_{11}-0_{00}$, $2_{20}-2_{11}$, and $4_{22}-4_{13}$), as well as their far-IR (FIR) dust continuum. %
By combining our data with previous measurements from the literature we estimate dust masses and temperatures, continuum optical depths, IR luminosities, and the star-formation rates from the FIR continuum. {\rm We model the H$_{2}$O lines using the MOLPOP-CEP radiative transfer code and find that water vapor lines in our quasar host galaxies are primarily excited in warm dense (gas kinetic temperature and density of $T_{\rm kin} = 50\,{\rm K}$, $n_{\rm H_{2}}\sim 10^{4.5}-10^{5}\,{\rm cm^{-3}}$) molecular medium with water vapor column density of $N_{\rm H_{2}O}\sim 2\times10^{17}-3\times10^{18}\,{\rm cm^{-3}}$.} %
High-$J$ H$_{2}$O lines are mainly radiatively pumped by the intense optically-thin {\rm far-IR} radiation field associated with a warm dust component with temperatures of $T_{\rm dust}\sim 80-190\,{\rm K}$ that account for $<5-10\%$ of the total dust mass. In the case of J2310+1855, our analysis points to a relatively high value of the continuum optical depth at $100\,{\rm \mu m}$ ($\tau_{100}\sim1$). Our results are in agreement with expectations based on the {\rm H$_{2}$O spectral line energy distribution} of local and high-$z$ ultra-luminous IR galaxies and active galactic nuclei (AGN). {\rm The analysis of the Boltzmann diagrams highlights the interplay between collisions and IR pumping in populating the high H$_{2}$O energy levels and allows us to directly compare the excitation conditions in the targeted quasar host galaxies.}
In addition, the observations enable us to sample the high-luminosity part of the H$_{2}$O--total-IR (TIR) luminosity relations ($L_{\rm H_{2}O}-L_{\rm TIR}$). Overall, our results point to supralinear trends suggesting that H$_{2}$O--TIR relations are likely driven by IR pumping rather than the mere co-spatiality between the FIR-continuum- and the line-emitting regions. The observed $L_{\rm H_{2}O}/L_{\rm TIR}$ ratios in our $z>6$ quasars do not show any strong deviations with respect to those measured in star-forming galaxies and AGN at lower redshift. This supports the idea that H$_{2}$O can be likely used to trace the star-formation buried deep within the dense molecular clouds.}


               
   \titlerunning{Unveiling the warm dense ISM phase in $z>6$ QSOs}
   \authorrunning{Pensabene et al.}
   \maketitle
   
%

\section{Introduction}
\label{sect:introduction}
Quasars (or QSOs) at $z>6$ are among the most luminous sources in the early Universe within the first billion years after the Big Bang. Their active galactic nuclei (AGN) are fueled by a rapid accretion of matter ($>10\,M_{\astrosun}\,{\rm yr^{-1}}$) onto a central supermassive black hole (BH; $\Mbh\apprge10^{8}\,M_{\astrosun}$; see e.g., \citealt{Jiang+2007, deRosa+2011, deRosa+2014, Mazzucchelli+2017, Schindler+2020, YangJ+2021}) while in their host galaxies the impetuos consumption of huge gas reservoirs turns gas into stars at high rates (${\rm SFR}>100\,M_{\astrosun}\,{\rm yr^{-1}}$; see e.g., \citealt{Bertoldi+2003b, Bertoldi+2003a, Walter+2003, Walter+2009, Venemans+2018, Venemans+2020}). The star formation is often enshrouded in large amount of dust ($M_{\rm dust}\sim 10^{7}-10^{8}\,M_{\astrosun}$) which makes the host galaxies of $z>6$ quasars very luminous in (far-)infrared (FIR) wavelengths ($L_{\rm FIR}\sim10^{12}-10^{13}\,L_{\astrosun}$) typical of the brightest nearby ultra-luminous IR galaxies \citep[ULIRGs; see, e.g.,][]{Decarli+2018, Venemans+2018, Venemans+2020}. Quasars cannot be merely considered as a rare dramatic phenomenon but rather as a fundamental phase in the formation and evolution of massive galaxies. For this reason, characterizing the quasar host galaxies at the highest redshifts is key to get insights on the interplay between star formation and BH accretion in shaping the galaxies from cosmic dawn to the contemporary Universe. 

The last decades have witnessed a real revolution for the study of massive galaxies down to the Epoch of Reionization. The advent of sensitive facilities in the (sub-)mm bands, such as ALMA (Atacama Large Millimeter Array), and the recently upgraded IRAM/NOEMA (Institute de Radio Astronomie Millim\'etrique/Northern Extended Millimeter Array) have enabled astronomers to image the host galaxies of $z>6$ quasars with unprecedented details. Bright tracers of the interstellar medium (ISM) such as the fine-structure line (FSL) of the singly-ionized carbon \cii{}, or carbon monoxide (CO) rotational lines at mid-/low-$J$ ($J\apprle7$), have been widely targeted to study the cold dense gas ($T_{\rm gas}<100\,{\rm K}$, $n_{\rm H_{2}}\sim 10^{3}\,{\rm cm^{-3}}$) morphology and kinematics \citep[e.g.,][]{Walter+2004, Maiolino+2012, Wang+2013, Cicone+2015, Jones+2017, Shao+2017, Feruglio+2018, Decarli+2019, Neeleman+2019, Neeleman+2021, Venemans+2019, Venemans+2020, WangF+2019, Pensabene+2020, Walter+2022} in $z>6$ quasars. The combination of FSLs from atomic neutral and ionized gas (e.g., neutral carbon [C\RNum{1}], [C\RNum{2}],  singly-ionized nitrogen [N\RNum{2}], neutral and doubly-ionized oxygen [O\RNum{1}], [O\RNum{3}]), and FIR transition from molecular gas phase (primarily CO rotational lines), enabled to dissect the physical properties of the multiphase ISM and to study its excitation conditions \citep[e.g.,][]{Walter+2003, Walter+2018, Riechers+2009, Gallerani+2014, Venemans+2017c, Venemans+2017a, Carniani+2019, Novak+2019, YangJ+2019, Herrera-Camus+2020, Li+2020b, Li+2020, Pensabene+2021, Meyer+2022}. However, despite these efforts, further investigations are needed to push toward a comprehensive view of the extreme conditions of the ISM in primeval quasar host galaxies. To this purpose other tracers of the ISM probing different gas phases need to be targeted. 

Among dense gas tracers, water vapor (H$_{2}$O) is the strongest molecular emitter after high-$J$ CO transitions in the ISM of IR-bright galaxies \citep[see, e.g.,][]{Omont+2013, Yang+2016}. H$_{2}$O lines are excited in shocked-heated regions, outflowing gas, and in the warm dense molecular gas \citep[$n_{\rm H_{2}}\apprge 10^5 - 10^6 {\rm cm^{-3}}$, $T\sim 50-100\,{\rm K}$; see,][]{Gonzalez-Alfonso+2010, Gonzalez-Alfonso+2012, Gonzalez-Alfonso+2013, vanderTak+2016, Liu+2017} where the star-formation ultimately occurs. Therefore, unlike other tracers traditionally used to study the dense gas (such as, e.g., high-$J$ CO, {$^{13}$C}O HCO$^{+}$, HCN, HNC), water vapor probes star-forming regions deeply buried in dust or heated gas in extreme environments of AGN. On the down side, the complexity of the H$_2$O radiative spectrum implies that the excitation mechanism and the physical conditions of the ISM cannot be derived on the basis of the detection {\rm of a single or just a few lines} \citep[see, e.g.,][]{Gonzalez-Alfonso+2014, Liu+2017}. Both collisions and absorption of resonant radiation contribute to the excitation of H$_{2}$O levels. In particular, IR radiative pumping is the main excitation mechanism of the high-$J$ H$_{2}$O lines \citep{Weiss+2010, Gonzalez-Alfonso+2014}. This implies that water vapor emission provides us with insights on both the physical properties of the warm dense medium, and on the IR radiation field. %
Interestingly, the intensity of H$_{2}$O lines is found to be nearly linearly proportional with the total IR (TIR) luminosity of the galaxy \citep{Omont+2013, Yang+2013, Yang+2016, Liu+2017}.

Ground-based studies of water vapor emission in nearby galaxies are limited by telluric atmospheric absorption and consequently have been restricted to radio-maser transitions and a few transitions in luminous IR galaxies \citep[e.g.,][]{Combes+1997, Menten+2008}. On the other hand, pioneering studies using the {\it Infrared Space Observatory} \citep{Kessler+1996}, and {\it Herschel} \citep{Pilbratt+2010} have detected H$_{2}$O emission (mainly in absorption) in local galaxies sometimes exhibiting P-Cygni line profiles unambiguously associated with massive molecular outflows \citep[e.g.,][]{Fischer+1999, Fischer+2010, Gonzalez-Alfonso+2004, Gonzalez-Alfonso+2008, Gonzalez-Alfonso+2010, Gonzalez-Alfonso+2012, Goicoechea+2005, vanderWerf+2010, Weiss+2010, Rangwala+2011, Sturm+2011, Kamenetzky+2012, Spinoglio+2012, Meijerink+2013, Pereira-Santaella+2013, Liu+2017, Imanishi+2021}. Water vapor emission have also been detected in starburst galaxies, Hy/ULIRGs, and quasars at higher redshifts \citep[$z\sim1-3$; e.g.,][]{Bradford+2011, Lis+2011, Omont+2011, Omont+2013, vanderWerf+2011, Combes+2012, Lupu+2012, Bothwell+2013, Yang+2013, Yang+2016, Yang+2019, Yang+2020, Jarugula+2019}, up to $z\apprge6-7$ \citep{Riechers+2013, Riechers+2017, Riechers+2021, Riechers+2022, Apostolovski+2019, Koptelova+2019, Li+2020, YangJ+2020, Jarugula+2021} where many FIR water transitions are redshifted in the ALMA and NOEMA bands. %
However, these sporadic water vapor detections primarily rely on one/two lines thus leaving the warm and dense molecular ISM phase at $z>6$ largely uncharted. 

Given their large amount of dust and extreme IR-luminosity that can significantly enhance the emission from water vapor, $z>6$ quasar host galaxies are therefore ideal sources to be targeted in their H$_{2}$O lines. This paper is focused on the characterization of the warm dense ISM and the local IR radiation field in three $z>6$ quasar host galaxies, J2310+1855, J1148+5251, and J0439+1634. For these objects, \cii{}, multiple CO lines, and other ISM probes (including sparse H$_{2}$O lines) as well as dust continuum were previously detected \citep{Bertoldi+2003b, Robson+2004, Walter+2004, Walter+2009nat, Maiolino+2005, Maiolino+2012, Beelen+2006, Riechers+2009, Wang+2011, Wang+2013, Leipski+2013, Gallerani+2014, Cicone+2015, Feruglio+2018, Carniani+2019, Hashimoto+2019b, Shao+2019, YangJ+2019, Li+2020b, Li+2020, Yue+2021, Meyer+2022}. Here we present NOEMA 2-mm band observations toward J2310+1855, J1148+5251, and J0439+1634, targeted in their four ortho-/para-H$_{2}$O rotational lines ($3_{12}-3_{03}$, $1_{11}-0_{00}$, $2_{20}-2_{11}$, and $4_{22}-4_{13}$), together with the underlying FIR dust continuum. The targeted lines cover a wide range in H$_{2}$O energy levels associated to different gas regimes. After combining information from such tracers, in conjunction with data retrieved from the literature, we performed radiative transfer analysis by employing the {\sc MOLPOP-CEP} code \citep{AsensioRamos+2018}. This enabled us to study the H$_{2}$O excitation mechanisms and to put constraints on the physical properties of the warm dense phase of the ISM and the local IR dust radiation field in quasar host galaxies at cosmic dawn. 

This paper is organized as follows: in Sect.~\ref{sect:data_reduction} we present our quasar sample, the NOEMA observations, and we describe the data processing. In Sect.~\ref{sect:h2o_excitation} we outline the characteristics of H$_{2}$O emission lines and their excitation mechanism. In Sect.~\ref{sect:measurements} we describe the analysis of the calibrated data, and we report the H$_{2}$O line and FIR continuum measurements toward our three quasar host galaxies. In Sect.~\ref{sect:dust_fir_properties} we focus on the dust properties inferred from the analysis of the FIR dust continuum. In Sect.~\ref{sect:radiative_transfer_analysis} we describe the setup of our radiative transfer models obtained with MOLPOP-CEP code. In Sect.~\ref{sect:results_discussion} we compare our measurements with other studies in the literature and we present and discuss our results obtained by modeling the H$_{2}$O lines. Finally, in Sect.~\ref{sect:summ_conc} we summarize the results and we draw our conclusions.

Throughout this paper we assume a standard $\Lambda\rm{CDM}$ cosmology with $H_0=67.7\,\si{km\,s^{-1}Mpc^{-1}}$, $\Omega_{m}=0.307$, $\Omega_{\Lambda}=1-\Omega_{m}$ from \citet{PlanckColl+2016}.

\section{Observations and data reduction}
\label{sect:data_reduction}
The goal of this work is to capitalize on multiple H$_{2}$O lines in order to unveil their excitation mechanisms and to characterize the warm dense molecular ISM and the local dust IR radiation field in a sample of $z>6$ IR-bright quasar host galaxies. For these purposes, the quasars J2310+1855 ($z=6.00$), J1148+5251 ($z=6.42$), and J0439+1634 ($z=6.52$) are ideal targets visible from the NOEMA site. They are among the IR-brightest of all known quasars at $z\apprge 6$. For this reason, on the basis of the observed (almost linear) correlation between H$_{2}$O and IR luminosities \citep{Yang+2013, Liu+2017}, they are expected to have bright H$_{2}$O lines. In addition, they have been previously detected in some H$_{2}$O lines (\citealt{YangJ+2019, Li+2020b}, and Riechers et al. in prep.; see Table~\ref{tbl:line_measurements}) that we use here to complement our analysis. We note that the J0439+1634 is a lensed source with a host galaxy magnification factor in the range 2.6-6.6 ($95\%$ confidence interval, see \citealt{Fan+2019, YangJ+2019}; but see also \citealt{Yue+2021} for a detailed analysis on high-angular resolution [C\RNum{2}] ALMA observations). However, as also reported by \citet{YangJ+2019}, all our observed H$_2$O lines in J0439+1634 are well fitted by a single Gaussian profile (see Sect.~\ref{sect:measurements}) thus suggesting that a contribution from differential lensing affecting the kinematics structure of the source is likely minor \citep[see, e.g.,][]{Rivera+2019, Yang+2019}. All the quantities reported in this work for J0439+1634 have to be intended as purely observed quantities (unless differently specified). In Table~\ref{tbl:sample_prop} we summarize the properties of our sample.

%
%
\begin{table}
\caption{The sample of quasars studied in this work.}       
\label{tbl:sample_prop}      
\centering      
\resizebox{\hsize}{!}{
\begin{tabular}{ l c c c c}     
\toprule\toprule

{Object ID} 				&Short name		& 	R.A 				& DEC. 					&$z_{\rm [C\RNum{2}]}$$^{(1)}$ \\
						&				&	(J2000.0)				& (J2000.0)				&						\\
\cmidrule(lr){1-5}
SDSS J231038.88+185519.7	&J2310+1855 	&$23\si{\degree}10\si{\arcmin}38\si{\arcsecond}.882$		&$18^{h}55^{m}19^{s}.700$	&$6.0031$\\
SDSS J114816.64+525150.3	&J1148+5251	&$11\si{\degree}48\si{\arcmin}16\si{\arcsecond}.652$		&$52^{h}51^{m}50^{s}.440$	&$6.4189$\\	
2MASS J04394708+1634160 	&J0439+1634	&$04\si{\degree}39\si{\arcmin}47\si{\arcsecond}.110$		&$16^{h}34^{m}15^{s}.820$	&$6.5188$\\
\bottomrule                 
\end{tabular}
}
\tablefoot{$^{(1)}$Redshift of the source estimated from \cii{} line \citep[see,][]{Maiolino+2005,Wang+2013,YangJ+2019}. }
\end{table}

We observed the targeted quasars using the IRAM/NOEMA interferometer in compact (C or D) array configuration (project ID: S19DL). The PolyFix band 2 receivers was tuned to secure the 1080 -- 1130 GHz and 1190 -- 1245 GHz rest-frame frequency windows in the lower (LSB) and upper side bands (USB) respectively.

The quasar J2310+1855 was observed in two tracks on October 12 and November 10, 2019. The blazar 3C454.3 was observed as phase and amplitude calibrator, while absolute flux scale and the bandpass were set by observing the MWC349 calibrator. The precipitable water vapor (PWV) column density was $8-10\,{\rm mm}$ in the October track, and $4-6\,{\rm mm}$ in the November one. The quasar J1148+5251 was observed on September 09, 2019. We observed the sources 3C84 and LKHA101 as flux and bandpass calibrators. The PWV during the observations was $2-6\,{\rm mm}$. Finally, J0439+1634 was observed on August 09, 13, and 14, 2019. The sources 3C84, 3C454.3, MWC349, and LKHA101 were used as flux and bandpass calibrators, while the radio-loud quasar 0446+112 acted as phase and amplitude calibrator. The PWV was $8-12\,{\rm mm}$ in the first visit, and $2-4\,{\rm mm}$ in the other tracks.

We used the \textsc{clic} software in the \textsc{GILDAS} suite (February 2020 version) to reduce and calibrate the data. The high PWV and consequently high system temperature ($>200\,{\rm K}$ at the tuning frequencies) led to poor phase rms residuals in the August 09, 13 and October 12 tracks, which were thus flagged out. The calibrated cubes have $8726$, $9090$, and $6690$ visibilities for J2310+1855, J1148+5251 and J0439+1634 respectively, corresponding to $3.0$, $3.2$, and $2.3\,{\rm hr}$ on source (9 antennas equivalent).

We adopted the \textsc{mapping} software in the \textsc{GILDAS} suite to invert the visibilities and image the cube. The half power primary beam width is $29.2''$, $31.0''$, and $31.4''$ at the tuning frequencies of the three quasars ($172.443\,{\rm GHz}$, $162.776\,{\rm GHz}$, and $160.612\,{\rm GHz}$ for J2310+1855, J1148+5251, and J0439+1634, respectively). Using natural weighting of the visibilities, we obtain synthesized beams of $1.47\arcsec\times 0.79\arcsec$, $2.45\arcsec\times 2.13\arcsec$, and $2.81\arcsec\times 2.51\arcsec$, respectively. We resampled the cubes adopting a channel width of $50\,{\rm km\,s^{-1}}$. The achieved median RMS per channel in the LSB and USB respectively are $0.61\,{\rm mJy\,beam^{-1}}$ and $1.50\,{\rm mJy\,beam^{-1}}$ for J2310+1855, $0.42\,{\rm mJy\,beam^{-1}}$ and $0.60\,{\rm mJy\,beam^{-1}}$ for J1148+5251, and $0.52\,{\rm mJy\,beam^{-1}}$, $0.66\,{\rm mJy\,beam^{-1}}$ in the case of J0439+1634. 
\begin{figure}[!t]
	\centering
	\resizebox{\hsize}{!}{
	\includegraphics{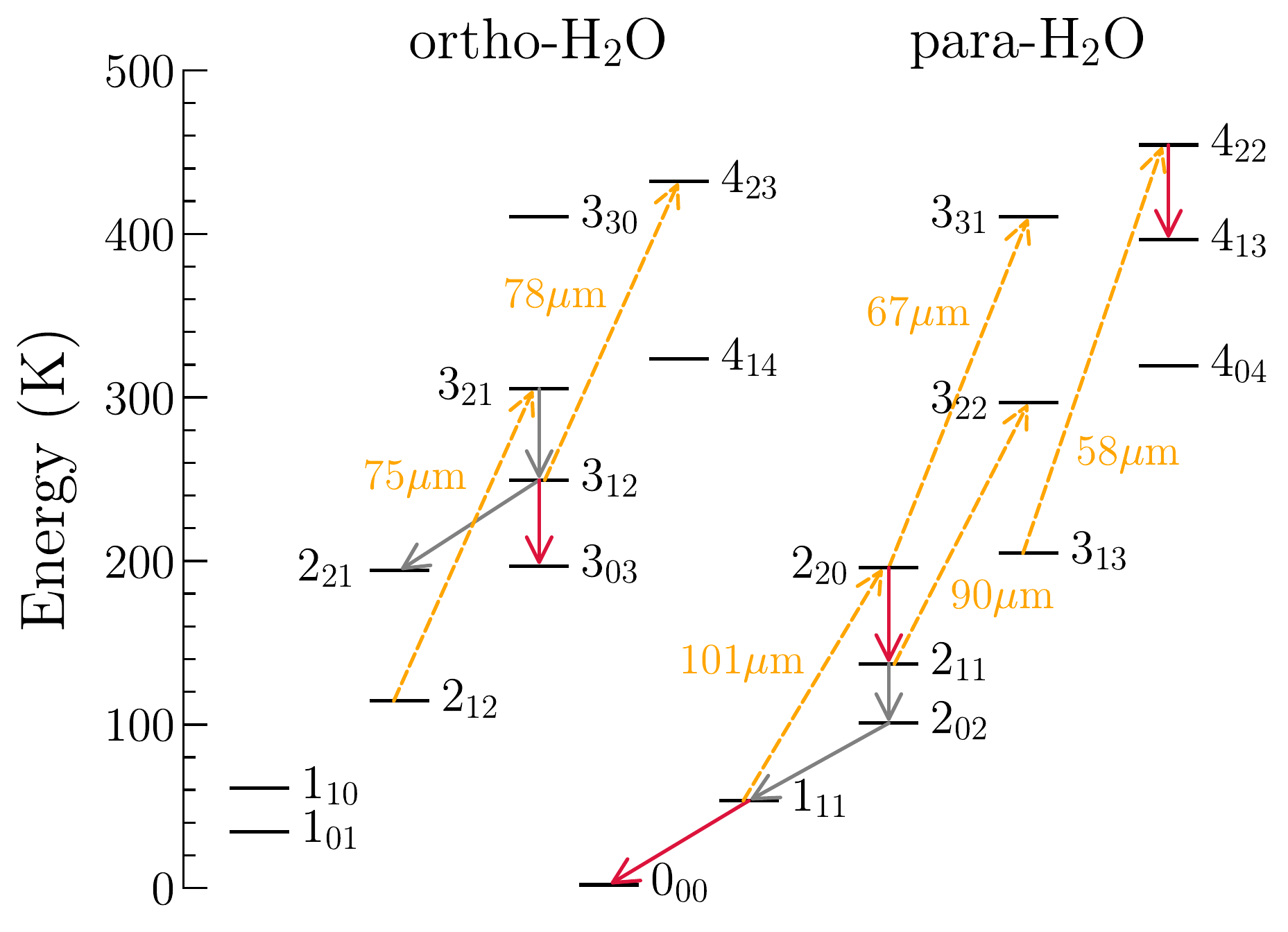}}
	\caption{Energy level diagram of H$_2$O. Downward solid red and gray arrows are the transitions reported in this paper and those available in the literature (see \citealt{YangJ+2019}, \citealt{Li+2020}, Riechers et al., in prep.). The upward dashed orange arrows indicate the FIR H$_2$O pumping (absorption) lines of interest. The respective wavelengths are also reported.}
         \label{fig:h2o_levels}
\end{figure}
\section{The H$_2$O emission lines and their excitation}
\label{sect:h2o_excitation}
In Fig.~\ref{fig:h2o_levels} we show the ladder structure of H$_{2}$O molecule (limited to energy levels $<500\,{\rm K}$). In the figure we indicate the H$_2$O transitions targeted in our NOEMA program together with the additional lines reported in the literature for our three targeted quasars. Properties of the relevant transitions such as the energy of the upper levels ($E_{\rm up}$), the rest frequencies ($\nu_{\rm rest}$), and the Einstein $A$ coefficients for spontaneous emission, are summarized in Table~\ref{tbl:obs_line}.

The excitation of the water vapor molecule is very sensitive to the physical conditions of the line-emitting region. As revealed by previous studies \citep[e.g.,][]{Gonzalez-Alfonso+2012, Gonzalez-Alfonso+2014, Liu+2017}, low-level transitions ($E_{\rm up}<250\,{\rm K}$) arise in warm collisionally-excited gas with kinetic temperature of $T_{\rm kin}\sim 30-50\,{\rm K}$, and clumps density of $n_{\rm H}\apprge 10^5\,{\rm cm^{-3}}$ that drives the low level populations toward the Boltzmann distribution with excitation temperature equal to the gas kinetic temperature ($T_{\rm ex} \simeq T_{\rm kin}$). This may occur even in environments with molecular gas density well below the critical density %
($n_{\rm crit}\sim 10^7-10^9\,{\rm cm^{-3}}$, e.g., \citealt{Faure+2007}) due to the large optical depth of such H$_2$O lines and radiative trapping effect that lowers the effective density at which the lines appear to be thermalized \citep[see, e.g.,][]{Poelman+2007}. The high-lying H$_{2}$O lines require instead radiative excitation by far-IR photons from warm dust ($T_{\rm dust}\sim70-100\, {\rm K}$) that are then re-emitted through a fluorescence process. This is illustrated in Fig.~\ref{fig:h2o_levels} where we report five far-IR pumping transitions ($58\,{\rm \mu m},\,75\,{\rm \mu m},\,78\,{\rm \mu m},\,90\,{\rm \mu m},\,101\,{\rm \mu m}$) that account for the radiative excitation of some sub-mm lines. Interestingly, the low-excitation lines ($E_{\rm up}<150\,{\rm K}$) are predicted to become weaker or completely disappear under the continuum level for increasing the dust temperature \citep[e.g.,][]{Liu+2017}. In particular, the para-H$_2$O ground state transition $1_{11}-0_{00}$ is not involved in any IR-pumping cycle and its flux is predicted to be negligible in regions where the IR pumping dominates. Indeed, the upper level $1_{11}$ can be populated only by absorption of the line photon at $269\,{\rm \mu m}$, or by a collisional event. This implies that in absence of significant collisional excitation (i.e., {\rm low gas density}) the H$_2$O $1_{11}-0_{00}$ line will be mainly detected in absorption in case of significant $269\,{\rm \mu m}$ continuum opacity \citep[e.g.,][]{Gonzalez-Alfonso+2004, Rangwala+2011}. On the other hand, in {\rm warm dense} gas region with low continuum opacity, the ground state transition is expected to be detected in emission. In this case, the $1_{11}$ level will be significantly populated by collisions thus enhancing the IR pumping cycle by absorption of $101\,{\rm \mu m}$ photons and boosting the $J=2$ para-H$_2$O lines. 
%
%
\setcounter{magicrownumbers}{0}
\begin{table}
\caption{List of the targeted H$_2$O emission lines in quasars.}  
\label{tbl:obs_line}      
\centering      
\resizebox{0.8\hsize}{!}{
\begin{tabular}{ l c c c}     
\toprule\toprule
{Transition} & {$\nu_{\rm rest}$ $^{(1)}$} & ${ E_{\rm upper}}$ $^{(2)}$ & ${{\rm Log\,} A_{\rm ij}}$ $^{(3)}$ \\
		&	(GHz)			&	(K)				&	$({\rm s^{-1}})$	\\
\cmidrule(lr){1-4}
H$_2$O $1_{11}-0_{00}$	& $1113.343$ & $61.0$  		& $-1.73$\\ 
H$_2$O $2_{02}-1_{11}$	& $987.927$ & $100.8$		& $-2.23$  \\ 
H$_2$O $2_{11}-2_{02}$	& $752.033$ & $136.9$		& $-2.15$\\ 
H$_2$O $2_{20}-2_{11}$  & $1228.789$ & $195.9$ 	& $-0.58$\\
H$_2$O $3_{12}-3_{03}$  & $1097.365$ & $249.4$ 	& $-1.78$\\
H$_2$O $3_{12}-2_{21}$  & $1153.127$ & $249.4$ 	& $-2.57$\\
H$_2$O $3_{21}-3_{12}$  & $1162.912$ & $305.2$ 	& $-1.64$\\
H$_2$O $4_{22}-4_{13}$  & $1207.639$ & $454.3$ 	& $-1.55$\\
\bottomrule                 
\end{tabular}
}
\tablefoot{$^{(1)}$Rest frequency. $^{(2)}$Energy of the upper level of the transition in units of the Boltzmann constant.  $^{(3)}$Base-10 logarithm of the Einstein $A$ coefficient for spontaneous emission. {\rm Data are taken from \citet{JPL+1998}}.}
\end{table}


In our NOEMA program we detected up to three para-/ortho-H$_2$O lines in each quasars that we complemented with other $J=2$, and $J=3$ H$_2$O lines from the literature \citep[][Riechers et al. in prep.; see Table~\ref{tbl:line_measurements}]{YangJ+2019, Li+2020}. This dataset allows us to cover a wide range of energy of H$_2$O levels ($50\,{\rm K} < E_{\rm up} < 500\,{\rm K}$), thus enabling for the first time a comprehensive analysis of the H$_2$O energetics in these high-$z$ quasars.

\section{Line and continuum measurements}
\label{sect:measurements}
\begin{figure*}[!htbp]
	\centering
	\resizebox{0.99\hsize}{!}{
	\includegraphics{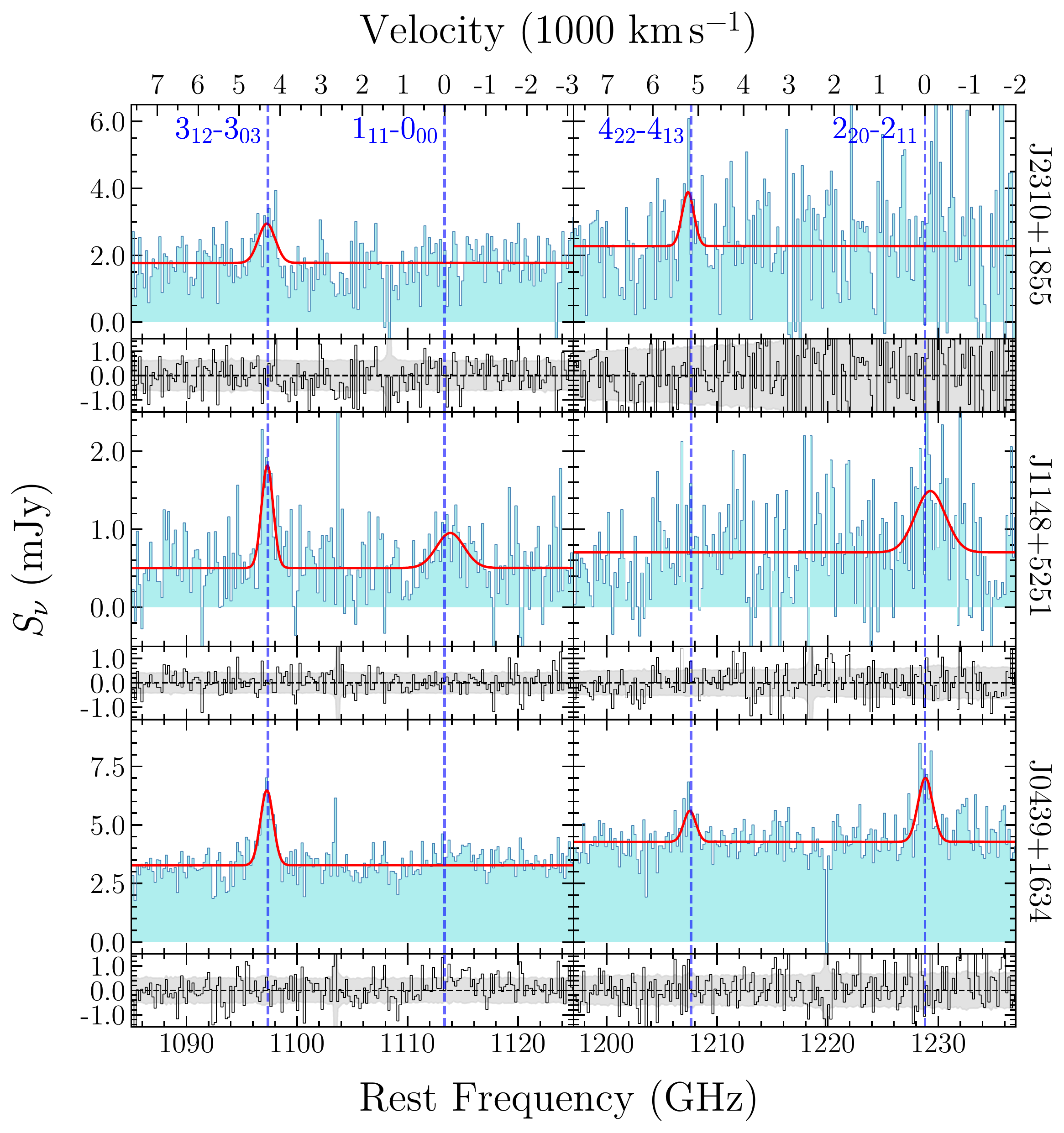}}
	\caption{NOEMA spectra of the three $z>6$ quasars J2310+1855, J1148+5251 and J0439+1634 (from top to bottom). We report the observed data in light blue and the best-fit models in red. The blue dashed lines indicate the rest frequencies of the targeted H$_2$O transitions (quantum numbers are also reported). At the bottom of each spectra we report the residuals (data-model). The gray-shaded areas show the noise RMS across the spectra.}
         \label{fig:spectra}
\end{figure*}

\begin{figure*}[!htbp]
	\centering
	\resizebox{\hsize}{!}{
	\includegraphics{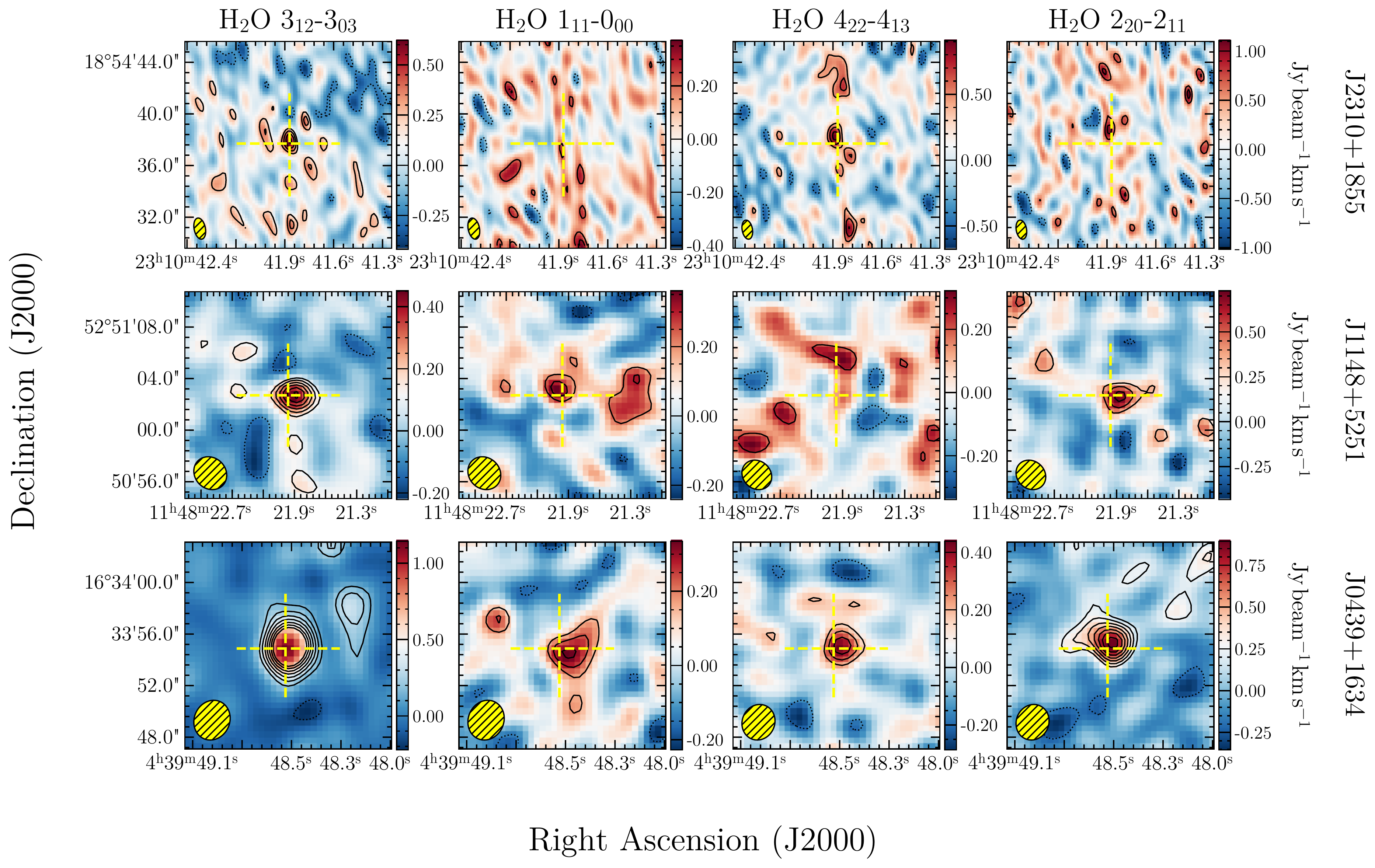}}
	\caption{The continuum-subtracted line velocity-integrated maps of the quasars J2310+1855, J1148+5251, J0439+1634 (from top to bottom). Yellow cross indicate the nominal coordinate of the source. The synthesized beam FWHM is reported at the bottom-left corners. The line maps are obtained by integrating over $1.4\times{\rm FWHM_{\rm line}}$ around the line central frequency. The black contours give the $[-4,-2,2,3,4,5,6,7,8,16,32]\sigma$ isophotes. Negative contours are shown as dotted lines.}
         \label{fig:line_maps}
\end{figure*}

Since our sources are not spatially resolved in the observations, we obtained the beam-integrated spectra by performing single-pixel extraction at the nominal coordinates of the targets (see Table~\ref{tbl:sample_prop}) from the data-cubes including the continuum emission. We then performed spectral fitting with a composite model including a single Gaussian component for the line and a constant for the continuum. We sampled the parameter space by using the package \texttt{emcee} \citep{Foreman+2013} that is a Markov chain Monte Carlo (MCMC) ensemble sampler developed in Python. This procedure allows us to effectively sample the posterior probability space. As data uncertainties in the likelihood estimates, we assume the 1-$\sigma$ Gaussian RMS of each imaged channel, and we neglect any systematic term. We therefore obtained the posterior probability distributions of the free parameters from which we derived the line and continuum measurements by computing the 50th percentile as the nominal values, and the 16th, 84th percentile as 1-$\sigma$ statistical uncertainties. We then derived the line luminosities as \citep[see, e.g.,][]{Solomon+1997}:
\begin{align}
\label{eq:lum}
&L_{\rm line}\,[L_{\astrosun}] = 1.04\times10^{-3}\,S\Delta\varv\,\nu_{\rm obs}D_L^2\,,\\
&L'_{\rm line}\,[{\rm K\,km\,s^{-1}pc^2}] = 3.25\times10^7\,S\Delta\varv\frac{D_L^2}{(1+z)^3\,\nu_{\rm obs}^2},
\label{eq:lum1}
\end{align}
where $S {\Delta \varv}$ is the velocity-integrated line flux in ${\rm Jy\,km\,s^{-1}}$, $\nu_{\rm obs}$ is the observed central frequency of the line in GHz, and $D_L$ is the luminosity distance in Mpc. The relation between Eq.~\ref{eq:lum} and ~\ref{eq:lum1} is $L_{\rm line}=3\times 10^{-11}\nu_{\rm rest}^3L'_{\rm line}$. We show the spectra of the sources with the best-fit models in Fig.~\ref{fig:spectra} and we report the derived quantities in Table~\ref{tbl:line_measurements} together with the additional H$_{2}$O line detections from the literature. %
In Fig.~\ref{fig:line_maps} we show the continuum-subtracted line velocity-integrated maps, the latter are obtained by averaging the continuum-subtracted cubes over $1.2\times{\rm FWHM}$ around the line. This velocity range is expected to maximize the S/N assuming an ideal Gaussian line profile and constant noise over the spectral channels. Throughout the text, we report the significance on the measured line fluxes in unit of $\sigma = \sqrt{\Delta \varv\,{\rm FWHM}} \times \left\langle{\rm RMS}\right \rangle$, where $\Delta\varv$ is the channel width ($50\,{\rm km\,s^{-1}}$), FWHM is the full width at half maximum of the best-fit Gaussian line model (in ${\rm km\,s^{-1}}$), and  the $\left\langle {\rm RMS}\right\rangle$ is the median RMS noise of the line cube (see Sect.~\ref{sect:data_reduction}).
{\rm In cases where line emission is not detected, we report $3\sigma$ upper limits assuming a source-specific FWHM value computed as the inverse-variance-weighted mean of the measured FWHM of the detected lines.}
%
%
%
\setcounter{magicrownumbers}{0}
\begin{table*}[!htbp]
\caption{Measurements and derived quantities of the NOEMA spectra of the quasars.}  
\label{tbl:line_measurements}      
\centering   
\resizebox{\hsize}{!}{
\begin{tabular}{ l c c c c c | l c c}     
\toprule\toprule
			\multicolumn{9}{c}{\bf \large J2310+1855}\\
			\vspace{-2mm}\\
Emission line  		&$z_{\rm line}$ 	&FWHM 			 &$S\Delta\varv$ 		&$L_{\rm line}$ 		&$L'_{\rm line}$    &{Dust continuum} $^{(1)}$			&{$S_{\nu}$} &Ref. $^{(2)}$\\
				&	     			&(${\rm km\,s^{-1}}$) &(${\rm Jy\,km\,s^{-1}}$) 	&($10^9L_{\astrosun}$)	&($10^{10}\,{\rm K\,km\,s^{-1}\,pc^2}$)	& 	&(mJy)	&\\ 
\cmidrule(lr){1-9}
${\rm H_{2}O}\;3_{12}-3_{03}$ 	&$6.0037^{+0.0018}_{-0.0017}$ &$494^{+140}_{-128}$ 	&$0.61^{+0.16}_{-0.15}$   &$0.35^{+0.09}_{-0.09}$  &$0.8^{+0.2}_{-0.2}$		&\multirow{2}{*}{$158.207$ GHz}	&\multirow{2}{*}{$1.77^{+0.04}_{-0.04}$}	&\multirow{4}{*}{This work}\\
${\rm H_{2}O}\;1_{11}-0_{00}$ 	&$-$ 					  &$-$ 				&${\rm <0.26}$ 				&${\rm <0.15}$ 				&${\rm <0.34}$			\\
${\rm H_{2}O}\;4_{22}-4_{13}$  &$6.0047^{+0.0013}_{-0.0022}$ &$314^{+328}_{-132}$ 	&$0.6^{+0.3}_{-0.2}$ 	&$0.35^{+0.16}_{-0.13}$   &$0.6^{+0.3}_{-0.2}$		&\multirow{2}{*}{$173.738$ GHz}	&\multirow{2}{*}{$2.27^{+0.08}_{-0.08}$}\\
${\rm H_{2}O}\;2_{20}-2_{11}$  &$-$ 						&$-$ 				&${\rm <0.62}$ 				&${\rm <0.40}$ 				&${\rm <0.67}$			\\
\cmidrule(lr){2-8}
${\rm H_{2}O}\;2_{02}-1_{11}$  &						&					&$0.70 \pm 0.05$  		&$0.36 \pm 0.03$ 	&$1.16\pm0.08$			&$141.070$ GHz				&$1.42 \pm 0.03$					&Li+20\\
${\rm H_{2}O}\;3_{12}-2_{21}$	&						&					&$0.53 \pm 0.17$  		&$0.32 \pm 0.10$ 	&$0.7 \pm 0.2$				&\multirow{2}{*}{$164.496$ GHz}	&\multirow{2}{*}{$2.33 \pm 0.11$} 		&\multirow{2}{*}{Riechers+}\\
${\rm H_{2}O}\;3_{21}-3_{12}$  &						&					&$1.25 \pm 0.17$  		&$0.76 \pm 0.10$ 	&$1.5 \pm 0.2$	\\

\cmidrule(lr){1-9}
			\multicolumn{9}{c}{\bf \large J1148+5251}\\
			\vspace{-2mm}\\
${\rm H_{2}O}\;3_{12}-3_{03}$ 	&$6.4192^{+0.0007}_{-0.0009}$  &$337^{+110}_{-77}$ 	&$0.47^{+0.10}_{-0.09}$   &$0.30^{+0.06}_{-0.06}$   &$0.70^{+0.15}_{-0.13}$ 	&\multirow{2}{*}{$148.510$ GHz} 	&\multirow{2}{*}{$0.50^{+0.03}_{-0.03}$}	&\multirow{4}{*}{This work}\\
${\rm H_{2}O}\;1_{11}-0_{00}$  	&$6.415^{+0.003}_{-0.004}$ 	   &$789^{+261}_{-256}$ 	&$0.36^{+0.14}_{-0.13}$ 	&$0.23^{+0.09}_{-0.09}$   &$0.52^{+0.20}_{-0.19}$ \\
${\rm H_{2}O}\;4_{22}-4_{13}$  &$-$ 					   &$-$ 				&${\rm <0.28}$ 				&${\rm <0.20}$ 				&${\rm <0.35}$				&\multirow{2}{*}{$164.202$ GHz}	&\multirow{2}{*}{$0.70^{+0.03}_{-0.03}$}\\
${\rm H_{2}O}\;2_{20}-2_{11}$  &$6.416^{+0.003}_{-0.003}$ 	   &$790^{+214}_{-223}$ 	&$0.64^{+0.20}_{-0.19}$ 	&$0.45^{+0.14}_{-0.14}$   &$0.8^{+0.2}_{-0.2}$ \\
\cmidrule(lr){2-8}
${\rm H_{2}O}\;2_{11}-2_{02}$	&			&								&$0.37 \pm 0.13$  		&$0.16 \pm 0.06$		 &$1.2 \pm 0.4$		&$101.367$ GHz				&$0.22 \pm 0.06$ 					&\multirow{4}{*}{Riechers+}\\
${\rm H_{2}O}\;2_{02}-1_{11}$  &			&								&$0.24 \pm 0.10$  		&$0.14 \pm 0.06$ 		 &$0.44 \pm 0.18$		&$133.164$ GHz				&$0.48 \pm 0.05$\\
${\rm H_{2}O}\;3_{12}-2_{21}$ 	&			&								&$0.33 \pm 0.11$  		&$0.22 \pm 0.07$		 &$0.44 \pm 0.15$		&\multirow{2}{*}{$155.277$ GHz}	&\multirow{2}{*}{$0.84\pm0.04$}\\
${\rm H_{2}O}\;3_{21}-3_{12}$ 	&			&								&$0.63 \pm 0.07$  		&$0.42 \pm 0.05$ 		 &$0.83 \pm 0.09$\\

\cmidrule(lr){1-9}
			\multicolumn{9}{c}{\bf \large J0439+1634}\\
			\vspace{-2mm}\\
${\rm H_{2}O}\;3_{12}-3_{03}$ 	&$6.5195^{+0.0004}_{-0.0004}$   &$372^{+44}_{-39}$ 	&$1.27^{+0.13}_{-0.12}$ 	&$0.82^{+0.08}_{-0.08}$   &$1.93^{+0.19}_{-0.19}$ 	&\multirow{2}{*}{$146.535$ GHz} 	&\multirow{2}{*}{$3.28^{+0.03}_{-0.03}$}	&\multirow{4}{*}{This work}\\
${\rm H_{2}O}\;1_{11}-0_{00}$  	&$-$ 					    &$-$ 				&${\rm <0.21}$ 				&${\rm <0.13}$ 				&${\rm <0.30}$ \\
${\rm H_{2}O}\;4_{22}-4_{13}$  &$6.5195^{+0.0010}_{-0.0010}$   &$321^{+169}_{-100}$  &$0.46^{+0.14}_{-0.12}$  &$0.32^{+0.10}_{-0.09}$    &$0.57^{+0.17}_{-0.15}$ 	&\multirow{2}{*}{$162.286$ GHz}	&\multirow{2}{*}{$4.27^{+0.04}_{-0.04}$}\\
${\rm H_{2}O}\;2_{20}-2_{11}$  &$6.5185^{+0.0006}_{-0.0006}$   &$382^{+82}_{-61}$ 	&$1.12^{+0.16}_{-0.15}$ 	&$0.81^{+0.12}_{-0.11}$    &$1.36^{+0.20}_{-0.18}$ \\
\cmidrule(lr){2-8}
${\rm H_{2}O}\;3_{12}-2_{21}$ 	&						   &					&$0.9 \pm 0.2$  		&$0.61 \pm 0.13$ 		&$1.2 \pm 0.3$			&\multirow{2}{*}{$154.667$ GHz}		&\multirow{2}{*}{$3.50\pm0.04$}			&\multirow{2}{*}{Yang+19}\\
${\rm H_{2}O}\;3_{21}-3_{12}$ 	&						   &					&$1.1 \pm 0.2$  		&$0.75 \pm 0.14$ 		&$1.5 \pm 0.3$	\\
\bottomrule
\end{tabular}
}
\tablefoot{$^{(1)}$Here we report the central frequency of the LSB and USB of each frequency setup of this work. For the literature data, the dust continuum reference frequency is set to one of the H$_2$O line expected frequency (based on the redshift measurement from \cii{} line, see \citealt{Decarli+2018, YangJ+2019}, and references therein) encompassed in the continuum frequency bandwidth. $^{(2)}$References: Yang+19 \citep{YangJ+2019}, Li+20 \citep{Li+2020}, Riechers+ (Riechers et al., in prep.).}
\end{table*}

%
\begin{figure*}[!htbp]
	\centering
	\resizebox{\hsize}{!}{
	\includegraphics{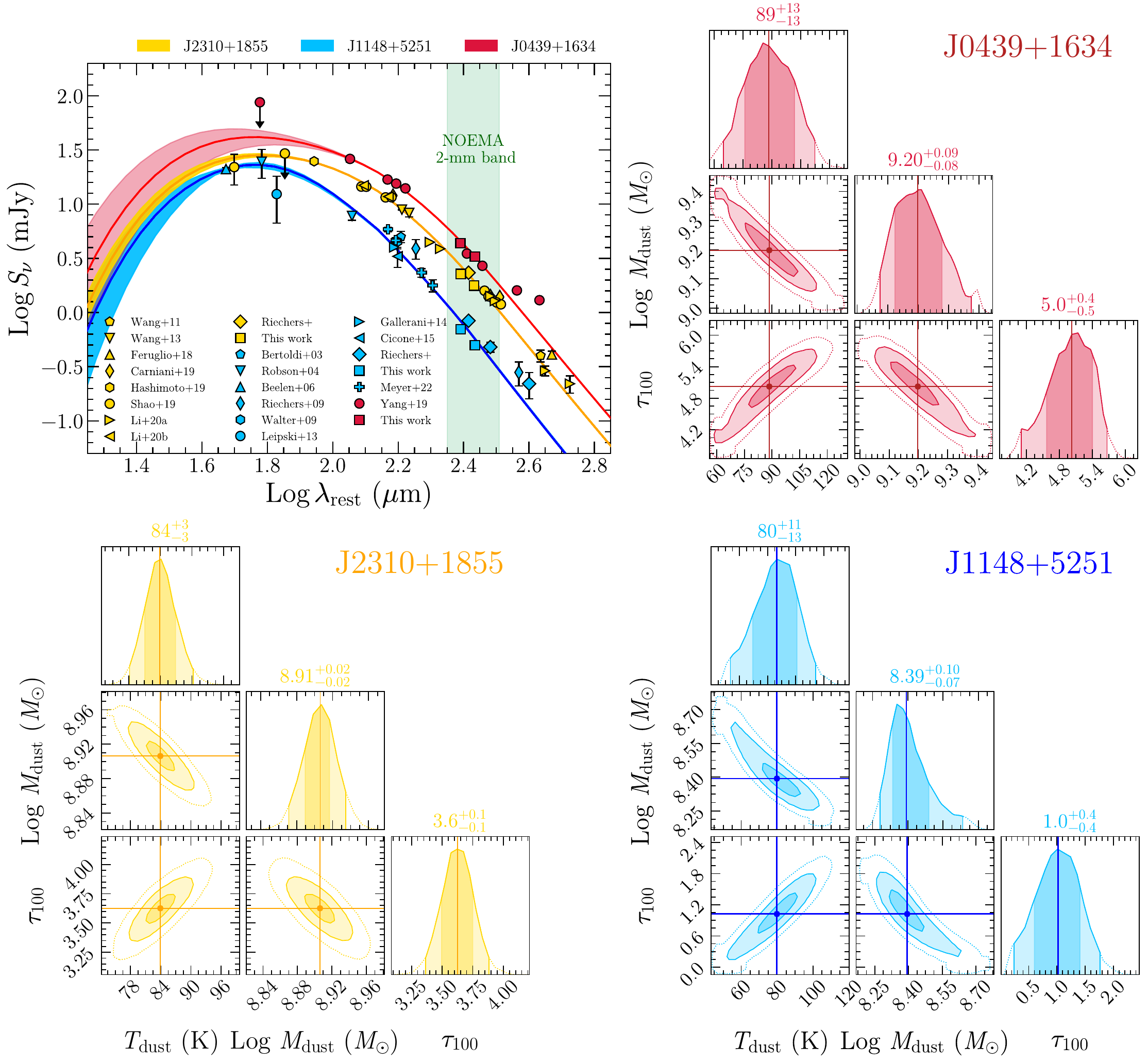}}
	\caption{Infrared dust continuum SED modeling of J2310+1855 (yellow), J1148+5251 (blue), and J0439+1634 (red). The upper-left panel shows continuum data retrieved from the literature {\rm derived for} the wavelength range $\lambda_{\rm rest}\approx 50-1000\,{\rm \mu m}$ and our NOEMA 2-mm continuum measurements (see the legend). The solid curves are the best-fit models to the observed data while shaded areas indicate the 1-$\sigma$ confidence intervals. The green band is the 2-mm wide-band of the NOEMA PolyFix correlator scaled to the quasar rest frame wavelengths using the average redshift of the quasars. The adjacent panels show the posterior probability distribution of the dust SED model (free) parameters. The best-fit values and uncertainties are reported at the top of the distributions and they are defined as the 50th, 16th, and 84th percentiles. The 2D contours show the $[1,2,3]\sigma$ confidence intervals that are also highlighted in the marginalized distributions. References: Bertoldi+03 \citep{Bertoldi+2003b}; Robson+04 \citep{Robson+2004}; Beelen+06 \citep{Beelen+2006}, Riechers+09,+ \citep[][Riechers et al. in prep.]{Riechers+2009};  Walter+09 \citep{Walter+2009nat}; Wang+11,+13 \citep{Wang+2011, Wang+2013}; Leipski+13 \citep{Leipski+2013}; Gallerani+14 \citep{Gallerani+2014}; Cicone+15 \citep{Cicone+2015}; Feruglio+18 \citep{Feruglio+2018}; Carniani+19 \citep{Carniani+2019}; Hashimoto+19 \citep{Hashimoto+2019b}; Shao+19 \citep{Shao+2019}; Yang+19 \citep{YangJ+2019}; Li+20a,+20b \citep{Li+2020b, Li+2020}; Meyer+22 \citep{Meyer+2022}.}
         \label{fig:dust_seds}
\end{figure*}
\section{Dust properties from FIR continuum}
\label{sect:dust_fir_properties}
The far-infrared continuum flux densities are key measurements to constrain the dust physical properties of galaxies. At (sub-)mm wavelengths the observed flux of $z>6$ quasar host galaxies is dominated by the dust re-emission of (rest-frame) UV photons from stars and the central AGN \citep[e.g.,][]{Beelen+2006, Leipski+2013, Leipski+2014}. The interstellar dust grains absorb and emit radiation with an efficiency that depends on the wavelength of the incident photons \citep{DraineLee1984} and therefore they do not behave as ideal blackbodies. The dust IR flux density at the rest-frame frequency $\nu$ emitted by a galaxy can be described by the dubbed ``graybody'' law: $S_\nu \propto (1-{\rm e}^{-\tau_\nu})B_\nu(T_{\rm dust})$, where $B_\nu(T_{\rm dust})$ is the Planck function depending on the dust temperature ($T_{\rm dust}$), and $\tau_\nu$ is the frequency-dependent dust optical depth. The observed flux density at frequency $\nu/(1+z)$ from a source at redshift $z$ can be expressed in terms of the dust mass ($M_{\rm dust}$) as \citep[see, e.g.,][]{Carniani+2019,Liang+2019}:
\eq{S_{\nu/(1+z)}=\frac{1-{\rm e}^{-\tau_\nu}}{\tau_\nu}\frac{(1+z)}{D_L^2}\kappa_\nu M_{\rm dust}B_\nu(T_{\rm dust}),
\label{eq:dust_flux}}
where $\kappa_\nu$ being the mass absorption coefficient, and $D_L$ is the redshift-dependent luminosity distance. The above equation provides us with a useful formula to estimate the dust mass and temperature from the observed (sub-)mm continuum flux densities. However, Eq.~\ref{eq:dust_flux} depends on largely unknown parameters such as the dust optical depth $\tau_\nu$, and the dust mass absorption coefficient $\kappa_\nu$ that are difficult to determine. Indeed, observational studies at low and high redshift showed that the interstellar dust emission becomes optically thick at wavelengths around $\lambda_{\rm rest} = 50-200\,{\rm \mu m}$ \citep[e.g.,][]{Blain+2003, Conley+2011, Rangwala+2011, Riechers+2013, Simpson+2017, Carniani+2019, Faisst+2020}. However, the strong contribution of AGN torus at MIR wavelengths prevents us from reliably sampling the Wien's tail of the dust SEDs in J2310+1855 and J1148+5251 \citep[see, e.g.,][]{Leipski+2014,Shao+2019}, and in the case of J0439+1634 we lack of continuum detections at (rest-frame) $<100\,{\rm \mu m}$ where the effect of dust optical depth is expected to be important. We therefore adopt the mass absorption coefficients for standard ISM from \citet{Draine2003}. We then extrapolated $\kappa_{\nu}$ at any frequency in the (rest-frame) range $6-1000\,{\rm \mu m}$ via a linear interpolation of the tabulated data, and we parametrize the continuum optical depth as $\tau_{\nu}=\tau_{100}\kappa_{\nu}/\kappa_{100}$, where $\tau_{100}$, and $\kappa_{100}$ are the continuum optical depth and the dust mass absorption coefficient at (rest-frame) $100\,{\rm \mu m}$.


The observed flux of Eq.~\ref{eq:dust_flux} must be corrected for the effect of the Cosmic Background radiation (CMB), the temperature of which is $T_{\rm CMB}(z=6)\approx 19\,{\rm K}$. In such conditions, the CMB acts a strong background source that attenuates the observed flux density, and provides a thermal bath heating the dust. \citet[][see also, \citealt{Zhang+2016}]{daCunha+2013} show that the recovered dust continuum observed against the CMB is given by
\begin{align}
&{S_{\nu/(1+z)}^{\rm obs}=\tonda{1-\frac{B_\nu(T_{\rm CMB}(z))}{B_\nu(T_{\rm dust}(z))}}S_{\nu/(1+z)}, \label{eq:flux_cmb_corr}}\\
&{T_{\rm dust}(z)=\graffa{\tonda{T_{\rm dust}^{z=0}}^{4+\beta}+\tonda{T_{\rm CMB}^{z=0}}^{4+\beta}\quadra{\tonda{1+z}^{4+\beta}-1}}^{1/(4+\beta)},
\label{eq:tdust_heated}
}
\end{align}
{\rm where $T^{z=0}_{\rm dust}$ and $T_{\rm CMB}^{z=0}$ are the corresponding dust and CMB temperatures at $z=0$, and $\beta$ is the dust spectral emissivity index}.  Throughout the rest of the paper we indicate with $T_{\rm dust}$ the \textit{actual} dust temperature (i.e., $T_{\rm dust}(z)$) including the contribution to the dust heating provided by the CMB as in Eq.~\ref{eq:tdust_heated}. The quasars J2310+1855, and J1148+5251 are widely studied in literature. For these sources, the dust physical properties are constrained using different SED decomposition models \citep[see, e.g.,][]{Leipski+2013, Leipski+2014, Shao+2019}. However, here we aim to obtain quasar dust properties by adopting {\rm the same} method for all the sources. For this purpose, we combine our continuum measurements in J2310+1855, J1148+5251, and J0439+1634 (see Table~\ref{tbl:line_measurements}) with the available continuum flux density measurements from the literature (see Fig.~\ref{fig:dust_seds}) in the range $\lambda_{\rm rest} \approx 50 -1000\,{\rm \mu m}$, where the dust emission is expected to be powered primarily by the re-processed optical/UV radiation of young stars in star-forming regions thus avoiding the MIR excess due to the AGN torus contribution and/or very hot dust components \citep{Casey+2012, Leipski+2013, Leipski+2014, Casey+2014}. {\rm \citet{Shao+2019} and \citet{Leipski+2014} show that the AGN in J2310+1855, and J1148+5251 starts to dominate the FIR flux at (rest-frame) wavelengths $<50\,{\rm \mu m}$ thus providing a significant contribution (up to $\sim 50\%$, depending on the details of the SED modeling) to the total IR luminosity. For the purpose of this work, we aim at obtaining the FIR luminosity contribution from the ISM dust emission only which is directly linked to the excitation of H$_{2}$O lines and which provide us with SFR estimate. This also allow us to perform a fair comparison with IR luminosities of other samples of star-forming galaxies and AGN.}

We therefore use Eq.~\ref{eq:dust_flux} and~\ref{eq:flux_cmb_corr} to fit the dust SEDs of our quasars by using $T_{\rm dust}$, $M_{\rm dust}$ and $\tau_{100}$ as free parameters. We explore the parameter space by adopting a Bayesian approach via the MCMC ensemble sampler \texttt{emcee} \citep{Foreman+2013}. In this procedure, we treated continuum data as independent measurements with Gaussian statistical uncertainties ignoring any systematics. We adopted shallow box-like priors on free parameters as follows: $T_{\rm CMB}(z)\le T_{\rm dust}\le150\,{\rm K}$, $6.0\le\log M_{\rm dust}\le10.0$, and $0.01\le\tau_{100}\le150$ in line with the typical values observed in high-$z$ star-forming galaxies and quasars \citep[e.g.,][]{Beelen+2006, Leipski+2013, Leipski+2014}. %
In the case of J0439+1634, the available data are limited to the Rayleigh-Jeans (RJ) tail of the dust SED, where the observed flux density is $\propto T_{\rm dust}M_{\rm dust}$ thus making these parameters degenerate. In addition, the lack of data at shorter wavelengths leaves $\tau_{100}$ unconstrained. In order to fit the dust SED of J0439+1634 we therefore included a  Gaussian prior on $T_{\rm dust}$ in the log-likelihood function of the form $-0.5\{(T_{\rm dust}-70\,{\rm K})/(15\,{\rm K})\}^2$, with $70\,{\rm K}$ being the typical dust temperature independently found by \citet{Carniani+2019} for J1148+5251 and J2310+1855 taking into account the dust optical depth. %
We show fit results in Fig.~\ref{fig:dust_seds}. {\rm We then derived rest-frame FIR luminosities ($42.5-122.5\,{\rm \mu m}$,  $40-400\,{\rm \mu m}$; see \citealt{Helou+1985,Helou+1988})}, and TIR ($8-1000\,{\rm \mu m}$ \citealt{Sanders+2003}) luminosities of the sources, by integrating the best-fit models in the corresponding frequency range. Finally, we inferred the SFR, using the local scaling relation from \citet{Murphy+2011}; ${\rm SFR_{IR}}(M_{\astrosun}\,{\rm yr^{-1}})=1.49\times10^{-10}\,L_{\rm TIR}/L_{\astrosun}$, under the hypothesis that the entire Balmer continuum (i.e., $912\,\AA <\lambda<3646\,\AA$) is absorbed and re-irradiated by the dust in the optically thin limit. Here, a Kroupa initial mass function (IMF; \citealt{Kroupa2001}) is implicitly assumed, having a slope of $-1.3$ for stellar masses between $0.1-0.5\,M_{\astrosun}$, and $-2.3$ for stellar masses ranging between $0.5-100\,M_{\astrosun}$. In Table~\ref{tbl:dust_data} we report all the derived quantities obtained with our dust continuum modeling. We note that, under our working assumptions, any possible contribution of the central AGN to the IR luminosity will result in biases on derived quantities, in particular an overestimation of IR luminosities and the ${\rm SFR_{\rm IR}}$. 

The estimated dust temperature in both quasar host galaxies J2310+1855 and J1148+5251 is higher than that derived by \citet{Shao+2019}, and \citet{Leipski+2013, Leipski+2014}, whose performed quasar SED decomposition in UV/optical-FIR range. Also, in the case of J1148+5251, we find a higher $T_{\rm dust}$ value compared to other works at FIR wavelengths \citep{Cicone+2015, Meyer+2022}. In general, our $T_{\rm dust}$ values are higher than typical dust temperature measured in high-$z$ quasar hosts \citep[e.g.,][]{Beelen+2006, Wang+2007, Wang+2008, Leipski+2013, Leipski+2014}. However, the aforementioned studies typically assume optical thin dust emission and $\kappa_{\nu}\propto\nu^{\beta}$, with the dust emissivity index $\beta=1.6$. On the other hand, we found values of $T_{\rm dust}$ and $M_{\rm dust}$ that are consistent within $2\sigma$ with those found by \citet{Carniani+2019} which took into account the effect of the continuum optical depth. Indeed, our results point to optical thick conditions at $<100\,{\rm\mu m}$ in all the cases. %
However, we note that the derived dust parameters from SED fitting depend on the adopted functional form as well as the broad-band photometry used in the fit. Indeed, a single-temperature modified-blackbody cannot account for the superimposed emission from multiple dust components in galaxy. In general, the contribution to the TIR luminosity in the RJ regime of the dust SED is expected to be dominated by the vast cold dust reservoir ($T\sim20\,{\rm K}$), while warmer dust components ($T\apprge 50-60\,{\rm K}$) boost the IR-luminosity at shorter wavelengths \citep[e.g.,][]{Dunne+2001, Farrah+2003, Casey+2012, Galametz+2012, Kirkpatrick+2012, Kirkpatrick+2015}. The luminosity on RJ regime is only linearly proportional to the dust temperature, while $L_{\rm FIR}$ scales much faster 
near to the peak of the dust SED. For this reason, the presence of a warmer dust component may significantly bias our single-component modeling toward higher dust temperatures, even if the warm dust is a small fraction of the total dust mass. However, the available continuum measurements of our quasars hosts do not allow us to measure the potential small secondary peak associated to the cold dust on the RJ tail of the SED. Therefore, our dust temperature estimates should be considered as {\it luminosity-weighted}, in contrast to the {\it mass-weighted} dust temperature physically associated to the bulk of the cold dust emission in galaxies (see, e.g., \citealt{Scoville+2016, Behrens+2018, Liang+2019, Faisst+2020, DiMascia+2021, Sommovigo+2021}; but see also \citealt{Harrington+2021} for further discussion). 

\section{Radiative transfer analysis}
\label{sect:radiative_transfer_analysis}
The observed line and continuum luminosities incorporate key information on the physical properties of the emitting regions. Such information can be extracted by a forward modeling of the observed quantities under a number of simplified assumptions on the geometry of the emitting region and on the atomic/molecular excitations and radiative transfer processes. 
In this work we adopt the publicly-available radiative transfer code MOLPOP-CEP\footnote{\rm The code is available at the following link: \url{https://github.com/aasensio/molpop-cep}} \citep{AsensioRamos+2018} in order to simulate the emission of H$_2$O lines from a molecular cloud under the effect of a dust radiation field. Compared to other methods, this code solves "exactly" (i.e., in principle, at any level level of accuracy) the radiative transfer equations for a multi-line problem through the coupled escape probability approach \citep[CEP,][]{Elitzur+2006} in the case of a one-dimensional plane-parallel slab of gas that can present arbitrary spatial variations of the physical conditions. This code divides the source in a set of zones in which the level population equations are derived from first principles and solved self-consistently including interactions with the transferred radiation and possibly an external radiation field. Therefore, the emergent line fluxes are predicted as a function of the depth into the line-emitting region that can be directly compared with the observations. This level of sophistication is necessary since we are dealing with radiative transfer under very optically thick conditions, both in the continuum and in the lines.

Within MOLPOP-CEP we studied the water vapor emission lines by setting up uniform slab models of molecular cloud impinged by an external radiation field produced by the dust. The parameters of interest for the molecular system are the density of molecular hydrogen ($n_{\rm H_2}$), the kinetic temperature of the gas ($T_{\rm kin}$), the water abundance ($X_{\rm H_2O}$), and the H$_2$O column density ($N_{\rm H_2O}$). We characterized the external dust radiation field by adopting a single modified black body model of the form $(1-e^{-\tau_\nu})B_\nu(T_{\rm dust})$ impinging the molecular medium from a side. The radiation field is therefore determined by the dust temperature ($T_{\rm dust}$), and the continuum optical depth at each frequency ($\tau_\nu$). In addition, we take into account the effect of CMB at $z\sim 6$ by inserting a blackbody radiation field with temperature of $T_{\rm CMB}=19.08\,{\rm K}$ illuminating the molecular slab from both sides.

In order to model the H$_2$O emission in our quasar host galaxies, we assumed fiducial values of ${\rm Log}\,n_{\rm H_2}\,({\rm cm^{-3}}) = 4.5$, $T_{\rm kin}=50\,{\rm K}$, and $X_{\rm H_2O}=2\times 10^{-6}$ \citep[see, e.g.,][]{Meijerink+2005, Gonzalez-Alfonso+2010, Liu+2017, vanDishoeck+2021}. We therefore generated a $16\times25$ grid of models with different dust radiation fields by varying the dust temperature in the range $T_{\rm dust}=[45,195]\,{\rm K}$ (with $10\,{\rm K}$ linear spacing), and continuum optical depth at $100\,{\rm \mu m}$ in the range $\tau_{100} =[0.01, 150]$ ($\sim0.17$ dex spacing). The continuum optical depth at every wavelength is then determined by a tabulation available within the code, corresponding to the properties of standard ISM dust \citep[see,][]{AsensioRamos+2018}. For para- and ortho-H$_2$O collisional excitations we assume both {\rm ortho- and para-H$_2$} molecules as collisional partners by adopting collisional rate coefficients from {\rm \citet{Daniel+2011}}. The radiation field impinges the molecular system and the code computes the radiation transfer all the way into the cloud until the water vapor column density reaches $N_{\rm H_2O}=1\times10^{19}\,{\rm cm^{-2}}$ ($N_{\rm H_2}=5\times10^{24}\,{\rm cm^{-2}}$). The allowed range of the various parameters are set in order to encompass the typical values observed in local and high$-z$ galaxies \citep[see, e.g.,][]{vanderWerf+2011,Gonzalez-Alfonso+2014, Yang+2016,Yang+2020, Liu+2017, Pensabene+2021}. Following the prescriptions reported in \citet{AsensioRamos+2018}, we set up the calculations by dividing the molecular cloud into 20 zones, achieving a relative accuracy in the solution of {\rm the non-linear level population equations of} $<0.01$ with the accelerated $\Lambda$-iteration method (ALI). 

The models adopted in this work are suitable to simulate the emission of a typical molecular cloud in a galaxy. The simplified assumptions make these models undoubtedly less adequate to provide a realistic picture of the complex ISM conditions in galaxies. In particular, our model is based on one-dimensional slabs with no defined volume. Therefore, our model cannot provide us with volume-integrated quantities (e.g., total IR luminosities, total line fluxes, molecular mass, etc.). We also note that a comprehensive modeling of water vapor emission should include the contribution of multiple gas and dust components in order to perform a fairer modeling of the collisionally-excited low-$J$ and radiatively-excited high-$J$ H$_{2}$O lines simultaneously (see, e.g., \citealt{Gonzalez-Alfonso+2010, vanderWerf+2011, Liu+2017, Yang+2020}; see, also, \citealt{Riechers+2022}). Specifically, a cold (extended) gas component associated to the collisional excitation of the low-lying H$_{2}$O lines mainly driven by the gas density and temperature, and (at least) one warm/hot (compact) component responsible for the radiative excitation of high-$J$ H$_{2}$O transitions which are mainly sensitive to the dust temperature. However, we verified that our current data do not enable us to constrain the gas temperature and density such that any multicomponent approach would be inconclusive. This drives our analysis strategy consisting in a single gas and dust component assuming fiducial values of the gas density, temperature, and water vapor abundance as discussed above.

\begin{figure*}[!htbp]
	\centering
	\resizebox{\hsize}{!}{
	\includegraphics{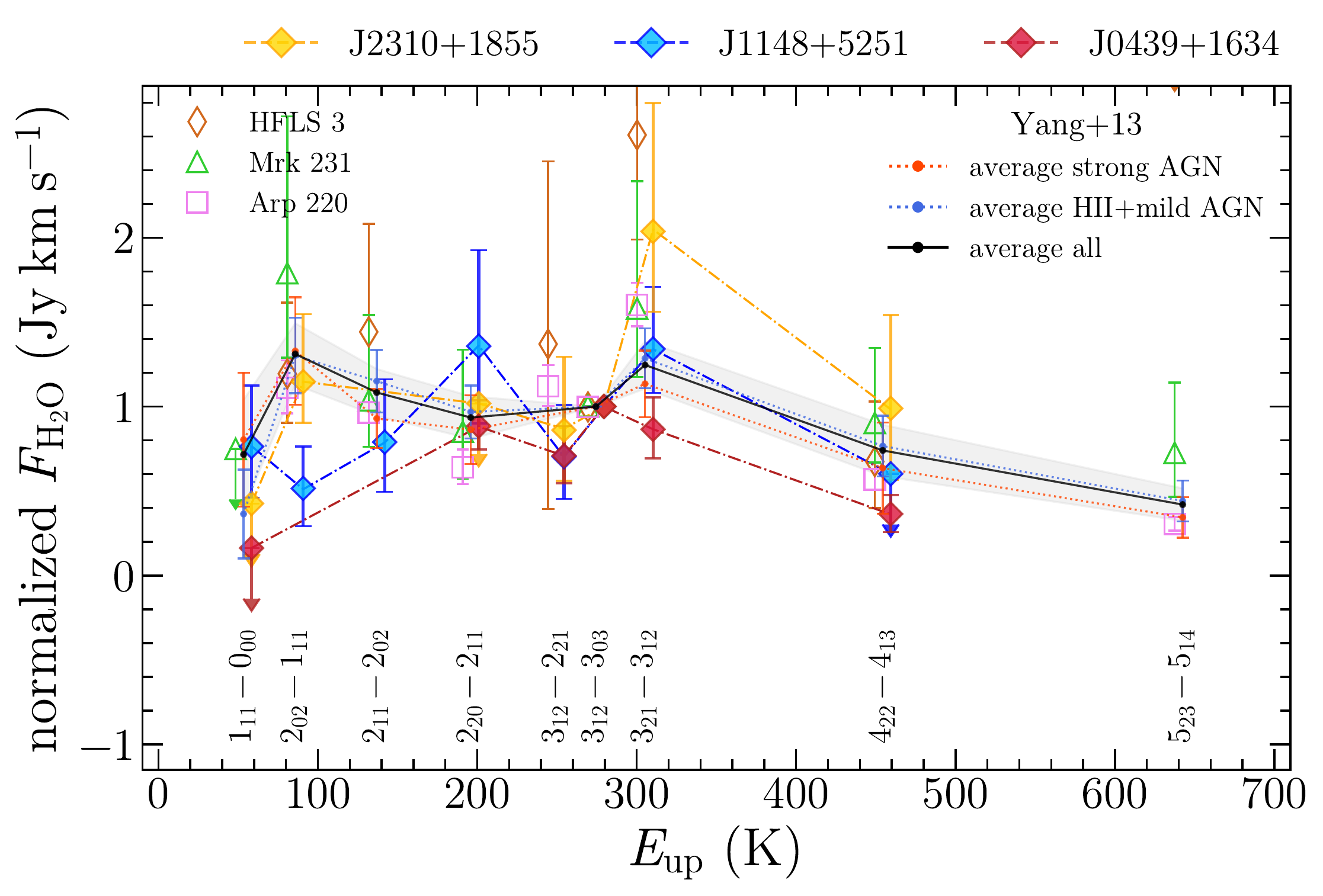}}
	\caption{H$_2$O ($3_{12}-3_{03}$)-normalized intensities (in ${\rm Jy\,km\,s^{-1}}$) as a function of the energy of the upper levels. Colored diamonds are the H$_2$O line ratios of our three quasars as indicated in the legend at the top of the panel. The red and blue dotted lines are, respectively, the average H$_2$O SLED of local strong-AGN- and HII+mild-AGN-dominated galaxies reported in \citet{Yang+2013}. The solid black line is the average SLED of the whole sample with the grey shadowed area representing the 1-$\sigma$ uncertainty. We also report data retrieved from the literature for the nearby ULIRG Arp 220 \citep{Rangwala+2011}, the AGN-dominated galaxy Mrk 231 \citep{Gonzalez-Alfonso+2010}, and the SMG HFLS 3 at $z=6.34$ \citep{Riechers+2013}. The energy of upper levels of $2_{02}-2_{11}$ and $3_{12}-3_{03}$ transitions were shifted for clarity to $-15\,{\rm K}$ and $+25\,{\rm K}$, respectively. We also slightly shifted the SLEDs of Arp 220, Mrk 231, HFLS 3 horizontally to $-5\,{\rm K}$, and those of our quasars to $+5\,{\rm K}$ for clarity.} 
         \label{fig:h2o_sled}
\end{figure*}
\setcounter{magicrownumbers}{0}
\begin{table}
\caption{Dust properties in quasar host galaxies.}  
\label{tbl:dust_data}
\centering      
\resizebox{\hsize}{!}{
\begin{tabular}{ l c c c}     
\toprule\toprule
{\rm Object ID}					&{\bf \large J2310+1855}	&{\bf \large J1148+5251}	&{\bf \large J0439+1634}\\
\cmidrule(lr){1-4}
$T_{\rm dust}$ (K)  							&$84^{+3}_{-3}$ 		&$80^{+11}_{-13}$		&$89^{+13}_{-13}$ ${^{(*)}}$  \\
Log $M_{\rm dust}/M_{\astrosun}$ 				&$8.91^{+0.02}_{-0.02}$	&$8.39^{+0.10}_{-0.07}$	&$9.20^{+0.09}_{-0.08}$ \\
$\tau_{100}$ 								&$3.6^{+0.1}_{-0.1}$		&$1.0^{+0.4}_{-0.4}$		&$5.0^{+0.4}_{-0.5}$ \\
$L_{\rm FIR[42.5-122.5\,\mu m]}\,(10^{13}\,L_{\astrosun})$ $^{(1)}$ 		&$1.87^{+0.06}_{-0.06}$		&$1.56^{+0.05}_{-0.06}$		&$3.1^{+0.5}_{-0.5}$ \\
$L_{\rm FIR[40-400\,\mu m]}\,(10^{13}\,L_{\astrosun})$ $^{(2)}$		&$2.08^{+0.06}_{-0.06}$		&$1.66^{+0.05}_{-0.06}$		&$3.5^{+0.5}_{-0.5}$ \\
$L_{\rm TIR}\,(10^{13}\,L_{\astrosun})$ $^{(3)}$ 	&$3.4^{+0.2}_{-0.3}$			&$2.7^{+0.3}_{-0.3}$		&$6.1^{+1.7}_{-2.3}$ \\
${\rm SFR_{TIR}}\,(M_{\astrosun}\,{\rm yr^{-1}})$	&$5071^{+335}_{-384}$		&$4062^{+474}_{-434}$		&$8985^{+2535}_{-3430}$ \\
\bottomrule                 
\end{tabular}
}
\tablefoot{${^{(1),}} {^{(2),}} {^{(3)}}${\rm Far-IR and total IR luminosities} obtained by integrating the best-fit modified black body model in the (rest-frame) wavelength range $42.5-122.5\,{\rm \mu m}$ \citep{Helou+1985}, $40-400\,{\rm \mu m}$ \citep{Helou+1988}, and $8-1000\,{\rm \mu m}$ \citep{Sanders+2003}, respectively. ${^{(*)}}$For this parameter we employed a prior in the fitting procedure (see text).}
\end{table}

\begin{figure*}[!htbp]
	\centering
	\resizebox{\hsize}{!}{
	\includegraphics{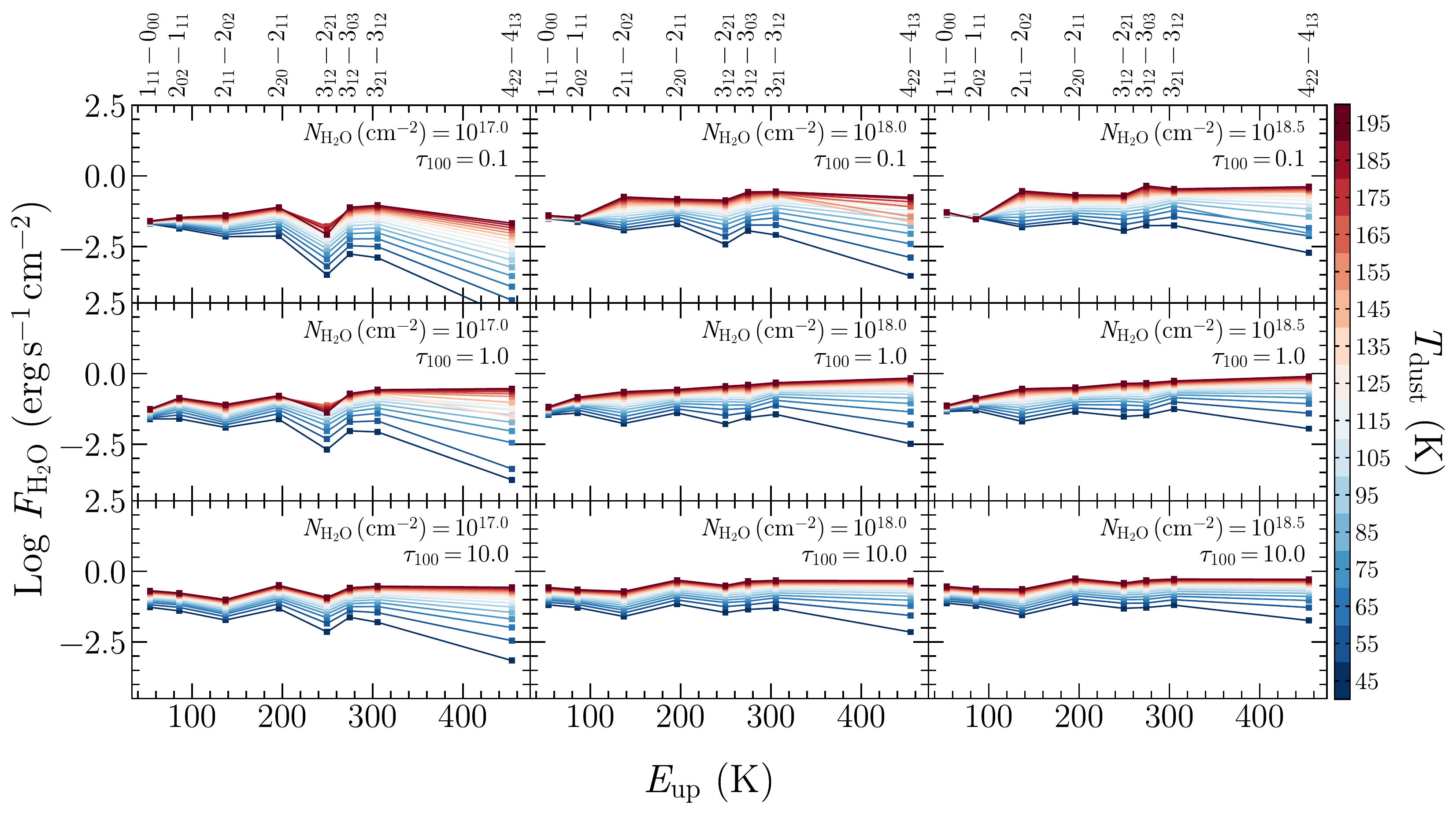}}
	\caption{H$_2$O line fluxes as a function of the energy of the upper levels obtained from our MOLPOP-CEP runs. Here we report output predictions for different values of water vapor column density (from left to right $N_{\rm H_2O}=10^{17}, 10^{18},10^{18.5}\,{\rm cm^{-2}}$) and continuum optical depth ({\rm from top to bottom ${\tau_{100}=0.1,1.0,10}$}) as reported at the upper right corner of each panel. The H$_2$O fluxes in each panel are color-coded according to the dust temperature ($T_{\rm dust}$).} 
         \label{fig:h2o_sled_models}
\end{figure*}

\section{Results and discussion}
\label{sect:results_discussion}
\subsection{H$_{\it 2}$O spectral line energy distributions}
\label{sect:h2o_sled}
In order to investigate the excitation of the H$_2$O lines we study the spectral line energy distribution (SLED), that is the water vapor line ratios as function of the energy of the upper level of transitions ($E_{\rm up}$). In Fig.~\ref{fig:h2o_sled} we compare the H$_2$O $(3_{12}-3_{03})$-normalized SLEDs (using line velocity-integrated fluxes in unit of ${\rm Jy\,km\,s^{-1}}$) in our quasars, with the average SLEDs of local ULIRGs, including cases with mild AGN contribution ("HII+mild AGN"), and AGN-dominated galaxies ("strong-AGN") as reported by \citet{Yang+2013}. In order to extend the comparison, in Fig.~\ref{fig:h2o_sled} we also report three detailed H$_2$O SLED available in the literature, for the local Type-1 Seyfert galaxy Mrk 231 \citep{Gonzalez-Alfonso+2010}, the local ULIRG Arp 220 \citep{Rangwala+2011}, and the SMG HFLS 3 at $z=6.34$ \citep{Riechers+2013}. To first order ({\rm within the measurement uncertainties}), our quasars show H$_{2}$O SLEDs that resemble the ones of other high-$z$ galaxies (with or without prominent AGN). 

More specifically, \citet{Yang+2013} report a high detection rate of H$_2$O $1_{11}-0_{00}$ in luminous AGN possibly indicating strong H$_2$O collisional excitation due to the high density gas in the AGN circumnuclear region (see discussion in Sect.~\ref{sect:h2o_excitation}). However, we detected the para-H$_2$O ground state transition only in quasar J1148+5251. In accordance with the discussion in Sect.~\ref{sect:h2o_excitation}, this may suggest that the bulk of the H$_{2}$O emission in this source arises from a warm ISM with low continuum opacity. %
At the same time, J1148+5251 is the only one source in our sample that is not detected in its H$_2$O $4_{22}-4_{13}$ transition. This might point to a weaker contribution of the warm dust in J1148+5251 compared to the other two sources. %

{\rm The H$_2$O $3_{21}-3_{12}$ is the brightest observed H$_{2}$O line in most of our quasar host galaxies. This is particularly evident for J2310+1855 that shows a peak in the H2O SLED at the position of this line as in the average local SLED of ULIRGs and AGN}. The H$_{2}$O $3_{21}$ level is indeed not only approximately thermalized by collisions in the warm medium, but also efficiently populated by absorption of $75\,{\rm \mu m}$ photons of the warm dust \citep[e.g.,][]{Liu+2017}. In addition, in optically thin conditions every de-excitation in H$_2$O $3_{21}-3_{12}$ line will be followed by a cascade either in the H$_2$O $3_{12}-3_{03}$ or $3_{12}-2_{21}$ transition ({\rm see Fig.~\ref{fig:h2o_levels}}), with relative intensities determined by the A-Einstein coefficient of spontaneous emission (see Table~\ref{tbl:obs_line}). In this situation, \citet{Gonzalez-Alfonso+2014} predicted a $3_{21}-3_{12}$-to-$3_{12}-3_{03}$ flux ratio of 1.16, well consistent with that we measure in J1148+5251 (${0.9^{+0.2}_{-0.2}}$), and J0439+1634 (${1.3^{+0.4}_{-0.3}}$). On the other hand, the higher value of this ratio in J2310+1855 suggests that larger line and/or continuum optical depth possibly decreases the strength of $3_{12}-3_{03}$ line relative to $3_{21}-3_{12}$ via absorption of $3_{12}-3_{03}$ photons re-emitted in the $3_{21}-2_{21}$ line, or significant IR pumping of the H$_2$O $4_{23}-3_{12}$ transition due to absorption of continuum photons at $78\,{\rm \mu m}$ (see Fig.~\ref{fig:h2o_levels}). 

In optically thin conditions the H$_2$O $2_{20}-2_{11}$, $2_{11}-2_{02}$, and $2_{02}-1_{11}$ lines powered via pumping by $101\,{\rm \mu m}$ photons form a closed loop and are expected to have approximately equal fluxes due to the statistical equilibrium \citep[see, e.g.,][]{Gonzalez-Alfonso+2014, Liu+2017}. However, high dust temperature and continuum optical depth increase the efficiency in the $3_{22}-2_{11}$ $90\,{\rm \mu m}$ radiatively-pumped transition, thus decreasing the $2_{11}-2_{02}$ line relative to the other transitions within the loop. In particular, since the H$_2$O $2_{02}-1_{11}$ transition is predicted to be easily excited by collisions in the warm medium \citep{Liu+2017}, the $2_{02}-1_{11}$-to-$2_{11}-2_{02}$ flux ratio is expected to be $\apprge 1$ for optical thick continuum and high H$_2$O column density ($N_{\rm H_2O}>10^{17}\,{\rm cm^{-2}}$, see, e.g., \citealt{Gonzalez-Alfonso+2014, Liu+2017}). In J1148+5251 we found that $2_{20}-2_{11}$, $2_{11}-2_{02}$ and $2_{02}-1_{11}$ exhibit consistent fluxes, with $2_{02}-1_{11}$-to-$2_{11}-2_{02}$ flux ratio $\sim 1$ thus suggesting optically thin conditions. Similarly, J2310+1855 shows a $2_{02}-1_{11}$-to-$2_{20}-2_{11}$ flux ratio lower limit ${\rm >0.9}$ possibly indicating very warm dust and optically thick continuum which make an efficient radiative pumping of the $3_{31}-2_{20}$ line due to absorption of $67\,{\rm \mu m}$ photons \citep[see, e.g.][]{Liu+2017}. We also note that the H$_2$O $2_{02}-1_{11}$ can be boosted relative to the $2_{20}-2_{11}$ due to the efficient collisional excitation of the lower line; however, the non-detections of the H$_2$O $1_{11}-0_{00}$ line in J2310+1855, the prominent peak in $3_{21}-3_{12}$, and the relative high flux in $4_{22}-4_{13}$ line point to a minor contribution of this effect in this source. On this line, J0439+1634 is expected to be an intermediate case in terms of excitation conditions with respect to the other two sources.

\begin{figure}[!htbp]
	\centering
	\resizebox{\hsize}{!}{
	\includegraphics{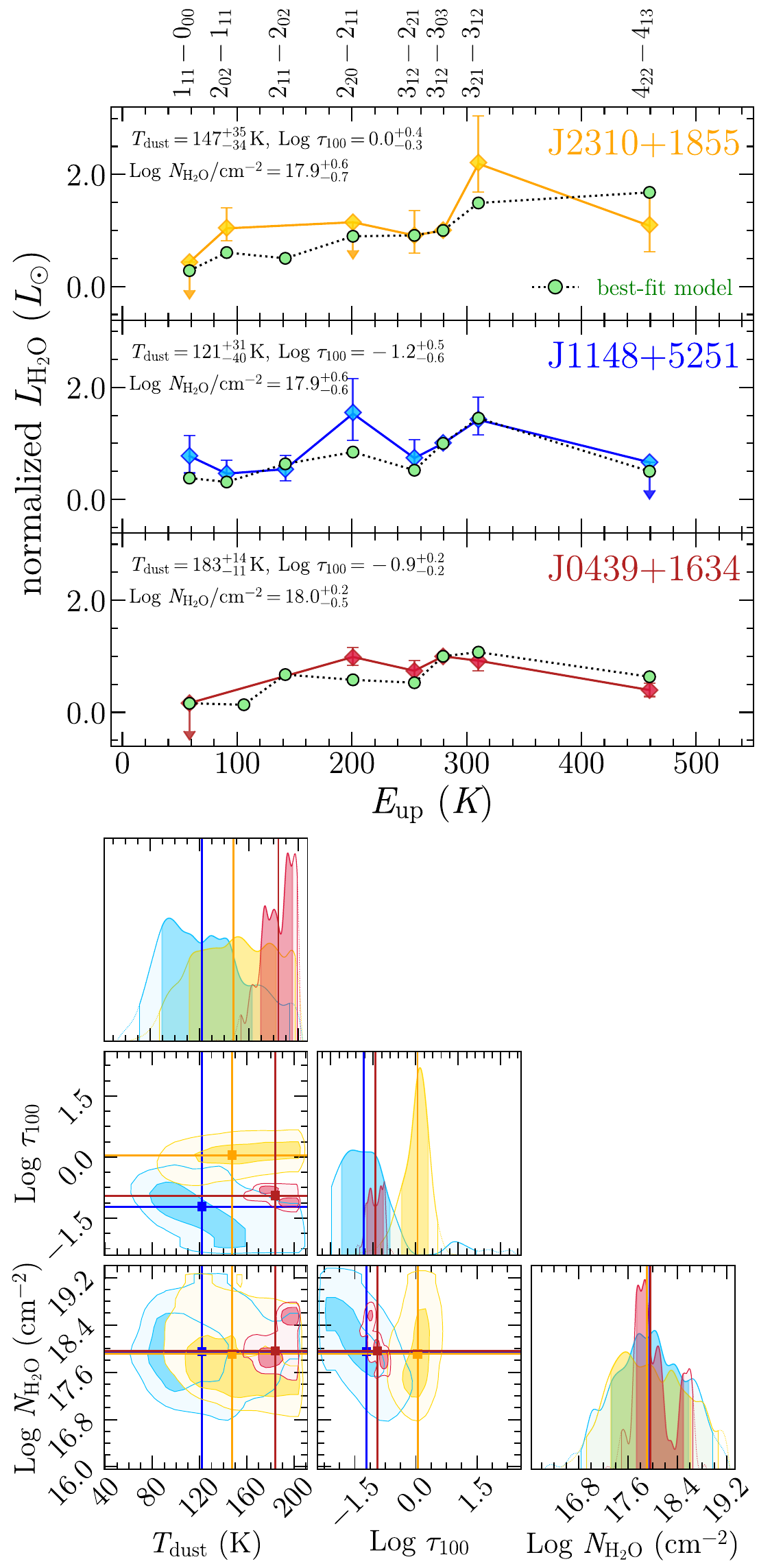}}
	\caption{Modeling of H$_2$O SLEDs of quasar host galaxies. {\it Top panels:} The colored diamonds are the observed H$_{2}$O line luminosities normalized to the H$_2$O $3_{12}-3_{03}$ line as a function of the energy of the upper level ($E_{\rm up}$). The green filled circles are the best-fit models obtained by using our MOLPOP-CEP grids. The best-fit parameters are reported at the top left corner of each panel. The reported uncertainties take into account only statistical errors ignoring any systematics. {\rm Both the $E_{\rm up}$ and the normalized $L_{\rm H_{2}O}$ are reported in linear scale.} {\it Bottom panels:} The MCMC output posterior probability distributions of free parameters. The contour plots show 1-$\sigma$, and $2\sigma$ confidence interval. Same intervals are reported in the marginalized distributions.} 
         \label{fig:h2o_sled_fits}
\end{figure}

\subsection{Modeling water vapor SLEDs}
\label{ssect:h2o_modeling}
In Fig.~\ref{fig:h2o_sled_models} we show our MOLPOP-CEP predictions of the H$_2$O flux for different values of $N_{\rm H_2O}$, $\tau_{100}$ and $T_{\rm dust}$. Our model predictions clearly reveal the effect of the IR-pumping of $J=3$, and $J=4$ lines. Indeed, for each couple of parameters ($N_{\rm H_{2}O}$, $\tau_{100}$), the fluxes of high-energy H$_{2}$O lines display a larger variation by increasing $T_{\rm dust}$ relatively to low-lying lines. Additionally, we found that H$_2$O fluxes at $E_{\rm up}>250\,{\rm K}$ systematically increase at high values of continuum optical depth, with larger absolute variations occurring at the largest $T_{\rm dust}$ values. This is not surprising as higher dust temperatures and continuum optical depths significantly boost the amount of IR photons that can be absorbed by H$_{2}$O molecules. This increases the fluxes of high-$J$ H$_{2}$O lines the levels populations of which are exclusively determined by radiative pumping. Such levels are expected to be described by a Boltzmann distribution with $T_{\rm ex}\sim T_{\rm dust}$ \citep[see, e.g.,][see also Sect.~\ref{sect:boltz_diagrams}]{Liu+2017}. In order to infer the ISM conditions in our quasar host galaxies, we employed our MOLPOP-CEP grids of models with which we perform fits of the observed quasar H$_2$O SLEDs. By assuming a single component model, we explore the (discrete) parameter space ($T_{\rm dust}$, $\tau_{100}$, $N_{\rm H_2O}$) adopting a Bayesian approach via the MCMC ensemble sampler \texttt{emcee} Python package \citep{Foreman+2013}. In this procedure, we assumed uniform priors for all the three free parameters within the ranges of the model grid. We therefore retrieved the posterior probability distributions by maximizing the log-likelihood function under the hypothesis that each data point follows an independent Gaussian distribution. We also took into account the measurement upper limits by assuming one third of their $3\sigma$ limits both as nominal values and uncertainties. In Fig.~\ref{fig:h2o_sled_fits} we show the best-fit H$_2$O SLED models and the posterior probability distributions of the free parameters. In Table~\ref{tbl:best-fit-sled} we report the best-fit parameters obtained from the fits of the H$_{2}$O SLEDs.

In general, our best-fit models reproduce well the observed H$_2$O SLEDs in our quasar host galaxies. All the modeled H$_2$O line luminosity ratios are consistent within the uncertainties with the measured data except for a few line ratios. The worst case is that of J2310+1855 SLED fit for which the best-fit model underestimates the H$_2$O $2_{02}-1_{11}$- and $3_{21}-3_{12}$-to-$3_{12}-3_{03}$ ratios, which however are consistent within $\sim2\sigma$ with the observations. However, we note that deviations in ratios involving low-$J$ lines may be expected if there is an additional low-excitation ISM component \citep[see, e.g.,][]{Gonzalez-Alfonso+2010, Liu+2017, Yang+2020}.

The posterior probability distributions of free parameters in Fig.~\ref{fig:h2o_sled_fits} point to very high dust temperature ranging in $T_{\rm dust}\sim 80-190\,{\rm K}$ with the highest values found in those sources in which the high-$J$ H$_2$O $4_{22}-4_{13}$ line is detected. This fact is in accordance with our discussion reported in Sect.~\ref{sect:h2o_sled}. The best-fit models predict optically thin continuum conditions at $100\,{\rm \mu m}$ (i.e., $\tau_{100} < 1$) except in the case of the J2310+1855 quasar for which ${\tau_{100}\sim 1}$ is favored within the uncertanties. The latter result is consistent {\rm within ${\sim 1.5\sigma}$} with what we found from the modeling of the dust SED at FIR wavelengths ($\tau_{100}\approx 3.6$). In the case of J1148+5251, the analysis of dust SED points to $\tau_{100} = 1.0\pm 0.4$ which is still consistent within $\sim2\sigma$ with the optically-thin regime suggested by the best-fit H$_{2}$O SLED model. Conversely, the dust SED modeling of J0439+1634 suggests instead optically thick conditions. However, the latter is affected by large uncertainties. Finally, high H$_2$O column density ${N_{\rm H_2O}\sim 2\times 10^{17}- 3\times10^{18}\,{\rm cm^{-2}}}$ is needed in order to match observations. {\rm Assuming our fiducial values of H$_{2}$O abundance ($X_{\rm H_{2}O}=2\times10^{-6}$) and gas density ($n_{\rm H_{2}}=10^{4.5}\,{\rm cm^{-3}}$), the best-fit H$_{2}$O column densities translate to molecular hydrogen column densities of $N_{\rm H_2}\sim 1\times10^{23}-2\times10^{24}\,{\rm cm^{-2}}$ which correspond to typical molecular cloud dimensions of $R=N_{\rm H_{2}}/n_{\rm H_{2}}\sim1-20\,{\rm pc}$. This sanity check suggests that our model results are reasonable\footnote{\rm We stress that these values has to be taken with caution since they depend on the model assumptions.}.} However, the best-fit results are affected by large uncertainties reflecting the spread of the parameter distributions in Fig.~\ref{fig:h2o_sled_fits}. The most precise predictions are obtained for J0439+1634, which is not surprising given the achieved high S/N of these observations. 
\setcounter{magicrownumbers}{0}
\begin{table}
\caption{Best-fit dust and gas properties of quasar host galaxies retrieved from H$_2$O SLED modelings.}  
\label{tbl:best-fit-sled}
\centering      
\resizebox{\hsize}{!}{
\begin{tabular}{ l c c c}     
\toprule\toprule
{\rm Object ID}					&{\bf \large J2310+1855}	&{\bf \large J1148+5251}	&{\bf \large J0439+1634}\\
\cmidrule(lr){1-4}
$T_{\rm dust}$ (K)  							&${147^{+35}_{-34}}$ 			&${121^{+31}_{-40}}$			&${183^{+14}_{-11}}$\\
Log $\tau_{100}$ 							&${0.0^{+0.4}_{-0.3}}$			&${-1.2^{+0.5}_{-0.6}}$			&${-0.9^{+0.2}_{-0.2}}$ \\
Log $N_{\rm H_2O}$ (${\rm cm^{-2}}$) 			&${17.9^{+0.6}_{-0.7}}$			&${17.9^{+0.6}_{-0.6}}$			&${18.0^{+0.2}_{-0.5}}$ \\
\cmidrule(lr){1-4}
$T_{\rm kin}$ (K)							&\multicolumn{3}{c}{$50$}\\
Log $n_{\rm H_2}$ (${\rm cm^{-3}}$)				&\multicolumn{3}{c}{$4.5$}\\
$X_{\rm H_2O}$							&\multicolumn{3}{c}{$2\times10^{-6}$}\\							
\bottomrule                 
\end{tabular}
}
\tablefoot{The first three row of the table are free parameters constrained in the H$_{2}$O SLED fit. {\rm The bottom part of the table summarizes the values for the other parameters adopted in our models and therefore valid for all the sources.}}
\end{table}


Our simple analysis reveals the presence of an intense warm dust component that can be responsible for the water vapor excitation in molecular clumps with high column density. In particular, for the quasar J2310+1855 we estimated a dust temperature of the optically-thick warm component of $T_{\rm dust}\sim150\,{\rm K}$, which yields a blackbody SED peak at around rest-frame $\approx 20\,{\rm \mu m}$. Remarkably, this value resembles the wavelength peak of the dusty torus component found by \citet{Shao+2019} who performed a dust SED decomposition of J2310+1855 continuum emission over a wide wavelength range. This result could indicate a significant MIR contribution of the dusty torus in triggering the water vapor emission, at least in the quasar nuclear region. However, we cannot reach a similar conclusion for the quasar J1148+5251 when comparing the best-fit dust radiation field temperature with the dusty-torus component inferred by \citet{Leipski+2013, Leipski+2014}. For this source, our radiative transfer analysis predict a dust temperature that is consistent within the uncertainties with that derived from dust SED modeling. Finally, we estimate a high dust temperature of $T_{\rm dust}\sim180\,{\rm K}$ in J0439+1634 suggesting the presence of a hot dust component in this quasar. Interestingly, \citet{Carniani+2019}, \citet{Yue+2021}, and \citet{Walter+2009nat, Meyer+2022} performed modeling of high-angular resolution ALMA/NOEMA observations of J2310+1855, J0439+1634, and J1148+5251, determining a FIR continuum half-light radius of $\approx 0.66\,{\rm kpc}$, $\approx 0.74\,{\rm kpc}$, and $\approx1.9\,{\rm kpc}$ respectively. This highlights that at least $50\%$ of the dust mass is contained in a compact central region, especially in the case of J2310+1855, thus further supporting a significant contribution of compact warm/hot dust component in heating the molecular gas in this quasar host galaxy. In any case, with the current analysis and data it is not possible to infer the geometrical properties (e.g., extension) of the bulk of the hot/warm dust components and the warm line-emitting region. Future high-angular resolution MIR/FIR studies will possibly reveal if the warm dust component we found could be actually associated to a central dusty torus in high-$z$ quasars. Overall, our results are in line with similar analysis conducted in other sources at low- and high-$z$ \citep[e.g.,][]{Gonzalez-Alfonso+2010, vanderWerf+2011, Liu+2017, Jarugula+2019, Yang+2020, Walter+2022}, predicting hot/warm dust components in the core regions of quasars, ULIRGs, SMGs and normal star-forming galaxies, notwithstanding the different modeling details. 

In order to find a constraint on the fraction of {\rm warm/hot} dust component in our quasar host galaxies we computed the expected observed flux density via Eq.~\ref{eq:dust_flux} using the best-fit parameters obtained from MOLPOP-CEP. In the case of J2310+1855, and J1148+5251, we scaled $M_{\rm dust}$ in order to match the {\it Herschel}/PACS \citep[Photodetector Array Camera Spectrometer;][]{Poglitsch+2010} data at (observed-frame) \mbox{$100\,{\rm \mu m}$ \citep{Leipski+2014, Shao+2019}}, while in the case of J0439+1634 we scaled the graybody model to match the JCMT/SCUBA-2 \citep[James Clerk Maxwell Telescope/Submillimetre Common-User Bolometer Array-2;][]{Holland+2013} $666\,{\rm GHz}$ (observed-frame) upper limit \citep{YangJ+2019}. As a result, we found that the fraction of warm/hot dust in our quasar host galaxies is $\apprle 5-10\%$ of the dust mass obtained from the dust FIR SED modeling. This result is in line with recent theoretical predictions on dust content in $z\sim6$ galaxies \citep{DiMascia+2021}.


\begin{figure}[!htbp]
	\centering
	\resizebox{\hsize}{!}{
	\includegraphics{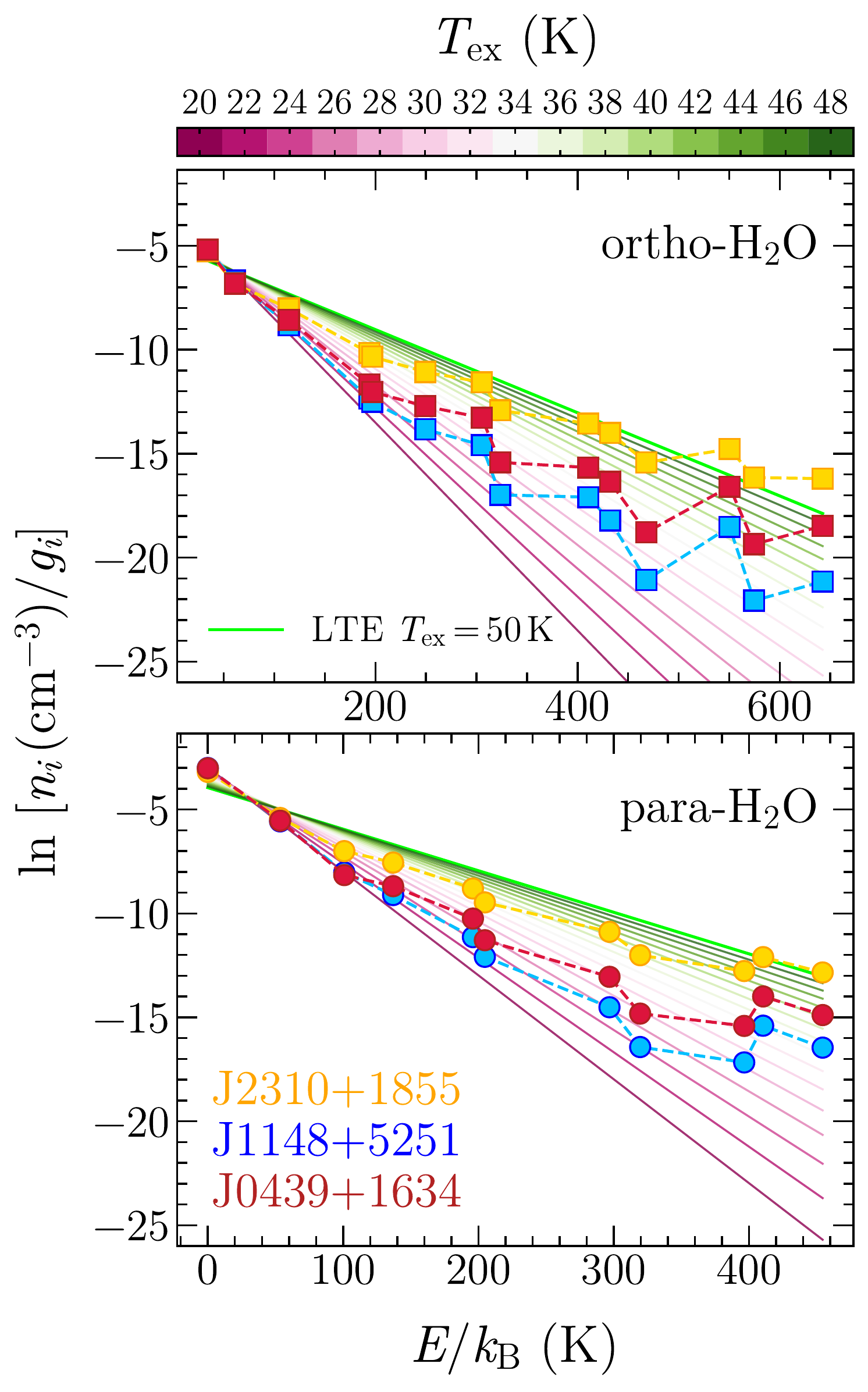}}
	\caption{Models of H$_2$O population level diagrams (upper and lower panel, respectively, for ortho- and para-H$_{2}$O) for our three quasar host galaxies. Each of the model {\rm shows the level volume density (weighted with their quantum degeneracy) at the $N_{\rm H_{2}O}$, $\tau_{100}$, and $T_{\rm dust}$ values corresponding to the MOLPOP-CEP best-fit H$_2$O SLED models (see, Sect.~\ref{ssect:h2o_modeling})}. The straight lines are the analytical population diagrams computed assuming Boltzmann distribution color-coded by the excitation temperature. The green line corresponds to the LTE case ($T_{\rm kin}=T_{\rm ex}$).}
    \label{fig:boltz_best-fit}
\end{figure}

\subsection{Boltzmann diagrams of H$_2$O levels}
\label{sect:boltz_diagrams}
We note that, the H$_2$O SLED, despite being a powerful diagnostic tool, remains quite elusive for a intuitive understanding of the ISM excitation conditions without a detailed investigation of the various line ratios. This is mainly due to the interplay of collisions and radiative pumping mechanism in exciting the water vapor lines. {\rm An analysis of the level population is required to better understand the H$_{2}$O excitation by varying the gas and dust radiation field properties . %
The Boltzmann statistics predict that in local thermodynamical equilibrium (LTE) conditions the level population of a specie obeys to $n_i/n_{\rm tot}=g_i/Q(T)\exp(-E_{i}/k_BT$), where $n_i$ is density of the {\it i-th} level at energy $E_{i}$, $g_{i}$ is its quantum degeneracy, $n_{\rm tot}$ is the total density of the species, and $Q(T)$ is the partition function. Therefore, a thermalized level population is represented by a straight line in a $\ln(n_i/g_i)-E_{i}/k_B$ diagram, the intercept and slope of which are determined by $n_{\rm tot}/Q(T)$, and $-1/T$, respectively. Collisional excitation is expected to drive the level population of low H$_{2}$O energy levels ($E< 100\,{\rm K}$) toward a Boltzmann distribution with an excitation temperature ($T_{\rm ex}$) near the gas kinetic temperature ($T_{\rm kin}$), while absorption of IR photons populates the high-energy level such that $T_{\rm ex}\sim T_{\rm dust}$ as soon as $T_{\rm dust}>T_{\rm kin}$. As a result of the interplay between collisional and radiative excitation, the H$_{2}$O level population can be approximated by a double $T_{\rm ex}$: one value for the low-$J$ lines (with $T_{\rm ex} \sim T_{\rm kin}$) and one for the higher-$J$ lines (with $T_{\rm ex}\sim T_{\rm dust}$).}

In Fig.~\ref{fig:boltz_best-fit} we show the population level diagrams for our three quasars as predicted by MOLPOP-CEP. {\rm These represent the volume densities of the level (weighted with their quantum degeneracy) at the H$_{2}$O column density, continuum optical depth, and dust temperature obtained via the H$_2$O SLED modeling discussed in the previous sections. In Fig.~\ref{fig:boltz_best-fit} we show a comparison of the population of the H$_{2}$O levels in our quasars.} The results show a significant contribution of the dust radiation field in exciting the high-$J$ H$_2$O lines in J2310+1855 compared to the other quasars. This is in line with the prominent peak of the H$_2$O SLED in the $3_{21}-3_{12}$ transition, and the mere detection of the H$_2$O $4_{22}-4_{13}$ line in the latter source. Indeed, significantly higher $T_{\rm dust}$ and $\tau_{100}$ values are inferred for this source. On the contrary, J1148+5251 appears much less excited in the high energy H$_2$O levels consistently with the lack of H$_2$O $4_{22}-4_{13}$ line emission and the detection of the ground state line $1_{11}-0_{00}$ indicating a minor contribution of the radiative pumping as a level populating mechanism. Finally, as also pointed out from the qualitative analysis in Sect.~\ref{sect:h2o_sled}, J0439+1634 represents an intermediate case between J1148+5251 and J2310+1855.

\subsection{$L_{\rm H_2O}-L_{\rm TIR}$ correlations}
Previous studies revealed the existence of nearly linear correlations between {\rm the luminosity of H$_{2}$O lines ($L_{\rm H_2O}$) and the total infrared luminosity ($L_{\rm TIR}$)} extending over $\sim12$ orders of magnitudes from the young stellar objects \citep[YSOs,][]{SanJose-Garcia+2015,SanJose-Garcia+2016}, where H$_2$O molecules are collisionally excited in shocked gas \citep[e.g.,][]{Mottram+2014}, to high-$z$ galaxies \citep{Omont+2013,Yang+2013, Yang+2016,Liu+2017}, in which radiative pumping plays an important role in populating the high-$J$ H$_{2}$O levels. However, these correlations have been interpreted differently in the Galactic and extragalactic context. {\rm In protostellar environments the H$_2$O emission is spatially compact and located either in the very proximity of the protostars or in their molecular outflows that happens in the earliest stages of star formation. Here, the FIR luminosity traces the material in which the stars are forming in the protostellar envelopes but it has no direct effect on the water vapor excitation \citep[see, e.g.,][]{vanDishoeck+2021}.} 
On the other hand, in the extragalactic context the $L_{\rm H_2O}-L_{\rm TIR}$ correlations are thought to be the direct consequence of the IR pumping. In this sense, the water vapor emission arises in molecular clouds that are not necessarily co-spatial with the star-formation activity. However, recent studies \citep[see, e.g.,][and references therein]{vanDishoeck+2021} showed that low- and mid-$J$ H$_2$O ($E_{\rm up} < 300\,{\rm K}$) line ratios do not significantly differ in Galactic and extragalactic environments. This suggests a common mechanism for water vapor excitation extending from individual Galactic YSOs to high-$z$ sources. That is, the H$_2$O emission is likely a good tracer of star formation activity buried in the protostellar envelopes. Therefore, H$_2$O traces proportionally the SFR as in the case of other dense gas tracers \citep[e.g., HCN,][]{Gao+2004b,Gao+2004a}. In particular, if a large fraction of the mass of the warm molecular ISM, where the bulk of the H$_2$O emission arises, is spatially-correlated with the physical regions where most of the FIR is generated, the nature of the $L_{\rm H_2O}-L_{\rm TIR}$ correlations could be easily explained as driven by the size of the emitting region \citep[see, e.g.,][]{Gonzalez-Alfonso+2014, Liu+2017}. Indeed, similar linear correlations are found for CO transitions which are collisionally excited in the molecular gas \citep[see, e.g.,][]{Greve+2014,Lu+2014,Liu+2015,Kamenetzky+2016, Yang+2017}. These linear correlations %
reflect the well-established fact that star-formation (traced by $L_{\rm TIR}$) occurs within the molecular ISM (which mass is traced by the CO line emission). On the other hand, if the radiative pumping drives the excitation of H$_2$O emission in high-$z$ galaxies, then one might expect that this affects to some extent the $L_{\rm H_2O}-L_{\rm TIR}$ relation. %
\begin{figure*}[!htbp]
	\centering
	\resizebox{\hsize}{!}{
	\includegraphics{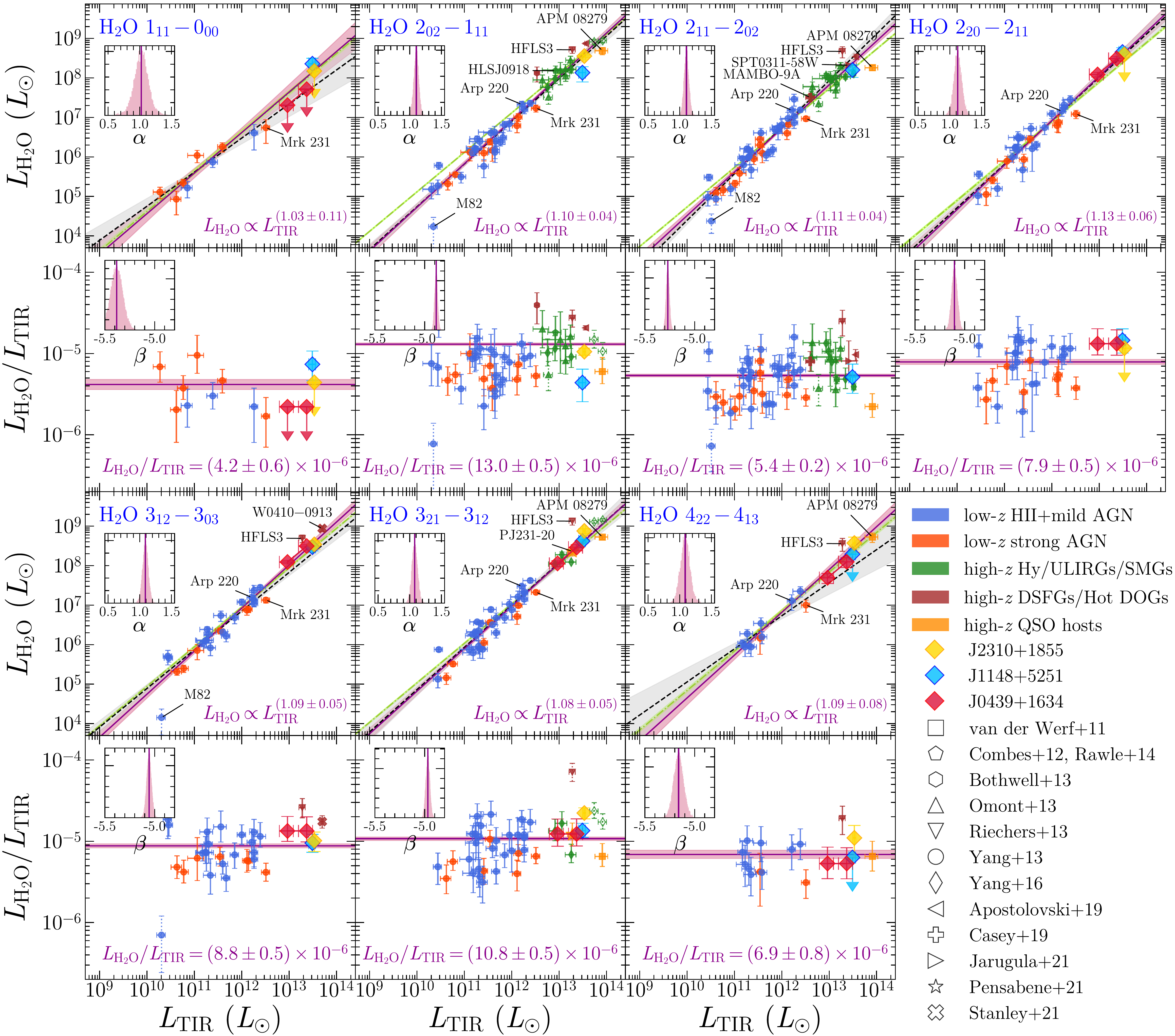}}
	\caption{Correlations between H$_2$O line luminosities and total IR luminosities. The reference transition is reported in the upper left corner in the upper panels. Lower panels report the correspondent $L_{\rm H_2O}/L_{\rm TIR}$ as a function of $L_{\rm TIR}$.  Data points show measurements of local and high-$z$ Hy/ULIRGs, SMGs, DSFGs/Hot DOGs and QSO hosts retrieved from the literature, color-coded by their type. Different symbols indicate the literature references according to the legend. For the lensed quasar J0439+1634 we report the $95\%$ confidence interval (red diamonds connected by a line) of intrinsic luminosities by adopting the magnification factor reported in \citet[][see also Sect.\ref{sect:measurements}]{YangJ+2019}. We use the mean magnification factor as the fiducial value in the fit. Data points with empty symbols are not corrected for gravitational lensing and were excluded from the fit. We also excluded those data reported with dotted error bars as explained in the text. Downward arrows are $3\sigma$ upper limits that we also ignored in the fit. The solid purple lines show our best-fit models. The inset panels show the posterior probability distribution of the slope $\alpha$ of the $L_{\rm H_{2}O}-L_{\rm TIR}$ correlations, and $\beta$ for $L_{\rm H_2O}/L_{\rm TIR}$. The best fit values are reported at the bottom right corner of each panel. For comparison we also show results obtained assuming an exact linear relation (dot-dashed green lines), and the best-fit models by \citet[][dashed black lines]{Yang+2013}. The shaded areas are 1-$\sigma$ confidence intervals.} 
         \label{fig:h2o_rel}
\end{figure*}
\citet{Gonzalez-Alfonso+2014} show that steeper than linear $L_{\rm H_2O}-L_{\rm TIR}$ relation is expected if (on average) $\tau_{100}$ is an increasing function of $L_{\rm TIR}$. By using FIR pumping models, they predict a $L_{\rm H_2O}\propto L_{\rm TIR}^{1.3}$ for the H$_2$O $2_{02}-1_{11}$ line and similar supralinear relations for the other H$_2$O transitions. However, \citet{Liu+2017} show that collisions significantly contribute to the excitation of the mid-$J$ H$_2$O lines, thus suggesting that the correlations are not the mere consequence of the radiative pumping effect. {\rm However, all the aforementioned investigations seem to suggest that $L_{\rm H_{2}O} - L_{\rm TIR}$ correlations are expected to be a consequence of either collisional excitation or IR pumping of H$_{2}$O transitions.} {\rm The lack of a clear correlation between H$_{2}$O and TIR luminosities would suggest that a larger-than-expected variety in the ISM properties (gas density, temperature, geometry, chemical composition, etc.) is in place in the emitting regions.}

In Fig.~\ref{fig:h2o_rel} we report the $L_{\rm H_2O}-L_{\rm TIR}$ correlations and the $L_{\rm H_2O}$-to-$L_{\rm TIR}$ ratios for the available lines in our quasar host galaxies. In order to study the correlations, we complement our data with previous observations from the literature. In particular, we retrieved measurements of local and high-$z$ Hy/ULIRGs samples \citep{Omont+2013,Yang+2013,Yang+2016}. We also include additional individual measurements of the quasar host galaxies APM 08279 at $z=3.9$ \citep{vanderWerf+2011}, and PJ231-20 (PSO J231.6576-20.8335) at $z=6.59$ \citep{Pensabene+2021}; the Hy/ULIRGs/SMGs HLSJ0918 \citep[HLSJ091828.6+514223,][]{Combes+2012, Rawle+2014} at $z=5.2$, and HFLS 3 at $z=6.34$ \citep{Riechers+2013}; the dusty star-forming galaxies (DSFGs) SPT 0538-50 at $z=2.78$ \citep[SPT-S J053816-5030.8,][]{Bothwell+2013}, MAMBO-9 at $z=5.85$ \citep[MM J100026.36+021527.9,][]{Casey+2019}, SPT 0346-52 at $z=5.66$ \citep[SPT-S J034640-5204.9,][]{Apostolovski+2019}, and SPT 0311-58 at $z=6.9$ \citep[SPT-S J031132-5823.4,][]{Jarugula+2021}; and the hot dust-obscured galaxy (Hot DOG) W0410-0913 \citep[][with $L_{\rm TIR}$ taken from \citealt{Fan+2016}]{Stanley+2021}. When possible, we report the values corrected for gravitational lensing.  Fig.~\ref{fig:h2o_rel} shows that our data provide new constraints to the ${\rm H_{2}O}-{\rm TIR}$ luminosity relations at $L_{\rm TIR}\apprge10^{13}\,L_{\astrosun}$, and $L_{\rm H_{2}O}\apprge 10^{8}\,L_{\astrosun}$ where only sparse detections are available in the literature thus far.

By assuming linear functional form of the type ${\rm Log}\,L_{\rm H_{2}O}=\alpha\,{\rm Log}\,L_{\rm TIR}+\beta$, we performed linear regressions in log-log space through a hierarchical Bayesian approach using the \texttt{linmix} Python package \citep{Kelly+2007}. In order to reduce possible biases in the fitting results, following \citet{Yang+2013, Yang+2016} we excluded %
M82 due to its peculiar very low values of its H$_2$O lines \citep[see,][for a discussion]{Weiss+2010,Yang+2013}, we then excluded SDP 81 due to missing flux filtered out by interferometer \citep{Omont+2013}, and the H$_2$O $3_{21}-3_{12}$ transition of HFLS 3 due to its high $L_{\rm H_2O}$-to-$L_{\rm TIR}$ ratio. Finally, we excluded all the sources not corrected for gravitational magnification and all upper limit measurements on H$_2$O luminosity. In the case of the lensed quasar J0439+1634 we derived the intrinsic luminosities by adopting a mean magnification factor of $4.6$ (see, Sect.~\ref{sect:measurements}, \citealt{YangJ+2019, Yue+2021}) and we used them as fiducial values in the fit. Finally, we also fit the $L_{\rm H_2O}/L_{\rm TIR}$ ratios by adopting the same aforementioned assumptions. Our best-fit results are reported in Fig.~\ref{fig:h2o_rel}, and Table~\ref{tbl:h2o_rel}. For comparison we report the relation found by \citep{Yang+2013} and the best-fit models obtained assuming an exact linear relation (i.e., $\alpha=1$). 

Overall, our best-fit slopes suggest slightly supralinear relations in all the cases except for the ground base transition H$_2$O $1_{11}-0_{00}$ that is consistent with a linear relation within the uncertainty. A similar result is valid in the case of H$_2$O $4_{22}-4_{13}$ line which relation is also roughly linear within $\sim$1-$\sigma$. {\rm However, the relations involving the H$_2$O $1_{11}-0_{00}$, and $4_{22}-4_{13}$ lines are constrained using a low number of data points available in the literature. This is reflected in the larger relative error on the correlation slopes with respect to the other lines.} The slopes of the correlations are all consistent within the uncertainties with those found by \cite{Yang+2013,Yang+2016}. %
The supralinear trend of the $L_{\rm H_2O}-L_{\rm TIR}$ correlations support the idea that, at least for mid- and high-$J$ H$_{2}$O lines, the correlations are likely driven by the IR pumping effect rather than the mere co-spatiality of the line- and IR-continuum-emitting region \citep[see,][]{Gonzalez-Alfonso+2014}. The intrinsic scatter about the ${\rm H_{2}O-TIR}$ relations are $\sim0.2\,{\rm dex}$ (see Table~\ref{tbl:h2o_rel}), comparable or even lower to that of the linear relation between CO lines and the FIR luminosity \citep[see, e.g.,][]{Liu+2015, Kamenetzky+2016}. Our analysis of the $L_{\rm H_2O}/L_{\rm TIR}$ ratios leads to similar average values than that reported by \citet{Yang+2013, Yang+2016}. We do not find any substantial difference in the H$_2$O-to-TIR luminosity ratios measured in our high-$z$ quasars than those of local AGN and star-forming galaxies, as expected if H$_{2}$O is a pure tracer of the buried star-formation activity. 

\setcounter{magicrownumbers}{0}
\begin{table}
\caption{Best-fit slopes of $L_{\rm H_2O}-L_{\rm TIR}$ relations and averaged $L_{\rm H_2O}/L_{\rm TIR}$ ratios.}
\label{tbl:h2o_rel}      
\centering      
\resizebox{\hsize}{!}{
\begin{tabular}{ l c c c }     
\toprule\toprule
{H$_2$O line} & 	{$\alpha$} 		&{$\hat{\sigma}\,$$^{(1)}$}						& $\left\langle L_{\rm H_2O}/L_{\rm TIR}\right\rangle\times 10^{-6}$ \\
\cmidrule(lr){1-4}
$1_{11}-0_{00}$	& $1.03\pm0.11$ 	&$0.23^{+0.10}_{-0.16}$				& $4.2\pm0.6$  	\\ 
$2_{02}-1_{11}$	& $1.10\pm0.04$ 	&$0.19^{+0.03}_{-0.04}$				& $13.0\pm0.5$	\\ 
$2_{11}-2_{02}$	& $1.11\pm0.04$ 	&$0.20^{+0.03}_{-0.03}$				& $5.4\pm0.2$		\\ 
$2_{20}-2_{11}$  	& $1.13\pm0.06$ 	&$0.18^{+0.05}_{-0.05}$				& $7.9\pm0.5$ 		\\
$3_{12}-3_{03}$  	& $1.09\pm0.05$ 	&$0.17^{+0.04}_{-0.05}$				& $8.8\pm0.5$ 		\\
$3_{21}-3_{12}$  	& $1.08\pm0.05$ 	&$0.20^{+0.03}_{-0.04}$				& $10.8\pm0.5$ 	\\
$4_{22}-4_{13}$  	& $1.09\pm0.08$ 	&$0.16^{+0.08}_{-0.09}$				& $6.9\pm0.8$ 		\\
\bottomrule                 
\end{tabular}
}
\tablefoot{$^{(1)}$The estimate of the dispersion of the intrinsic scatter about the regression line.}
\end{table}


%
%
%

%

\section{Summary and conclusions}
\label{sect:summ_conc}
We present NOEMA observations toward three IR-bright quasars (J2310+1855, J1148+5251, J0439+1634) at $z>6$, that were targeted in multiple water vapor lines as well as FIR continuum. These quasars were previously detected in multiple ISM probes including some H$_{2}$O lines. The lines targeted in the new observations are the para-/ortho-H$_{2}$O $3_{12}-3_{03}$, $1_{11}-0_{00}$, $2_{20}-2_{11}$, and $4_{22}-4_{13}$ emission lines, with most being detected in all three quasars. The combination of our new and the previous H$_{2}$O detections enable us to investigate the warm dense phase of the ISM, and the local IR dust radiation field in intensely star-forming galaxies at cosmic dawn. We pose new constraints on the $L_{\rm H_{2}O}-L_{\rm TIR}$ relations which enclose key information on the H$_{2}$O excitation mechanisms. In order to interpret our results and place quantitative constraints on the physical parameters of the ISM, we employed the MOLPOP-CEP radiative transfer code. The main results of this work are the following:
\begin{itemize}
\item We model the FIR dust continuum emission in all quasar host galaxies by assuming a single-component modified blackbody and taking into account the effect of the dust optical depth and the CMB contrast. With this approach we infer dust masses, temperatures, continuum optical depths, IR luminosities and the SFRs of the three quasar host galaxies (see, Sect.~\ref{sect:dust_fir_properties}, and Table~\ref{tbl:dust_data}). Our results suggest $T_{\rm dust}\sim80-90\,{\rm K}$ which is higher than the values previously reported in the literature assuming optically thin dust emission.
\item Our H$_{2}$O SLEDs do not show any obvious imprinting of powerful AGN activity characterizing our sources when compared to the SLEDs of local ULIRGs, AGN, as well as other high-$z$ sources. However, on the basis of the H$_{2}$O excitation physics, a detailed qualitative analysis of the individual line ratios suggests that the bulk of H$_{2}$O emission in J1148+5251 arises from warm ISM phase with a low continuum opacity. A similar analysis point to the presence of relatively warmer dust in J2310+1855, and J0439+1634 that boosts the excitation of high-$J$ H$_{2}$O lines compared to J1148+5251.
\item %
By using MOLPOP-CEP predictions we modeled the observed H$_{2}$O SLEDs of the quasar host galaxies. Our results reproduce the observed data well. The best-fit models reveal the presence of a warm dust component the temperature of which ranges in $T_{\rm dust}\sim 80-190\,{\rm K}$. Here, the highest temperatures are found in those sources that are detected in their high-lying $J=4$ H$_{2}$O transition. This warm dust component is presumably more compact than that probed by standard dust SED modeling. From our results we estimate that the mass fraction of the warm/hot dust component in our quasars host galaxies is $\apprle 5-10\%$.
\item {\rm Our best-fit models point to water vapor column densities in the range $N_{\rm H_{2}O}\sim2\times10^{17}-3\times10^{18}\,{\rm cm^{-2}}$. Optical thin conditions ($\tau_{100}\apprle1$) for the continuum are predicted in all sources except for J2310+1855 for which $\tau_{100}\sim 1$, consistent with the qualitative analysis of the H$_{2}$O SLEDs. Interestingly, our radiative transfer analysis on the H$_{2}$O SLED of J1148+5251, and J2310+1855 predicts values of $\tau_{100}$ which are consistent within $\sim2\sigma$ with the results obtained from the dust SED modeling. }
We find a significantly higher contribution of IR pumping in populating the high-$J$ H$_{2}$O lines in J2310+1855 compared to the other quasars. The excitation temperatures of $E>200\,{\rm K}$ H$_{2}$O levels are roughly $T_{\rm ex}\sim 40-50\,{\rm K}$, while they are around $T_{\rm ex}\sim25-30\,{\rm K}$ in J1148+5251.
\item We studied the correlations between H$_{2}$O and TIR luminosities and the H$_{2}$O-to-TIR luminosity ratios. Our observations, in combination with data from the literature, allow us to put constraints on the $L_{\rm TIR}\apprge10^{13}\,L_{\astrosun}$ and $L_{\rm H_{2}O}\apprge 10^{8}\, L_{\astrosun}$ part of the correlations. Our results suggest supralinear trends in all the cases except for the H$_{2}$O $1_{11}-0_{00}$ lines which correlation with $L_{\rm TIR}$ is well consistent with a linear trend, albeit being poorly sampled. Overall, our results support the idea that, at least for mid- and high-$J$ H$_{2}$O transitions, the $L_{\rm H_{2}O}-L_{\rm TIR}$ correlations are driven by the radiation pumping of the lines rather than the cospatiality of the H$_{2}$O-line- and IR-continuum-emitting regions. The analysis of the $L_{\rm H_{2}O}/L_{\rm TIR}$ ratios does not highlight any significant difference between values measured in high-$z$/local AGN, and that observed in local star-forming galaxies. Given the small intrinsic scatter of the H$_{2}$O-TIR relations, this result suggests that H$_{2}$O can be also used as a robust proxy of star-formation in high-$z$ quasar host galaxies.
\end{itemize}

We showed how the combination of multiple H$_{2}$O lines enables us to shed light on the properties of the warm molecular medium in massive galaxies in the Epoch of Reionization. Water vapor lines are also powerful diagnostics of the warm dust component that is difficult to unveil through a simple analysis of the dust continuum SED. However, at $z\sim 6$, many key H$_{2}$O lines are shifted in ALMA/NOEMA (sub-)mm bands such that they can be simultaneously targeted. In particular, at $z=6$, the combination of H$_{2}$O $3_{12}-3_{12}$, $3_{12}-2_{21}$ (the latter of which is blended with the CO($10-9$) line), or the H$_{2}$O $3_{12}-3_{03}$ together with the H$_{2}$O $1_{11}-0_{00}$, are encompassed in a $<3\,{\rm GHz}$ bandwidth and can be therefore secured with a single ALMA or NOEMA frequency setting. Despite the complexity of the water vapor excitation analysis that requires multiple H$_{2}$O line detections, our work suggests that the aforementioned combinations of lines maximize the scientific return of H$_{2}$O observations for the purpose of probing the warm dense ISM in galaxies at $z\apprge6$. Such observations will also provide relatively tight constraints on TIR luminosity (the H$_{2}$O $3_{21}-3_{03}-{\rm TIR}$ shows small intrinsic scatter). 

Our studies were limited by the low S/N of the current data. Further investigations on a larger sample of sources targeting multiple H$_{2}$O lines with deeper observations are required in order to determine if our results are characteristic of the $z>6$ quasar population. Overall, our analysis shows that the H$_{2}$O lines do not seem to be significantly affected by the extreme BH feedback in $z>6$ quasars when compared to water vapor emission in local star-forming galaxies. However, larger samples are required to further understand whether H$_{2}$O lines can provide information on AGN activity or if they can be used as pure tracers of star formation in high-$z$ quasar host galaxies. In addition, the brightness of H$_{2}$O lines compared to other typically used molecular tracers, makes them ideal probes of shocked medium by AGN-driven outflows. Future deeper (sub-)mm spectroscopic observations of primeval quasar host galaxies would also enable the systematic search for H$_{2}$O lines featuring P-Cygni profiles that are known to be a smoking gun of massive molecular outflows.
\begin{acknowledgements}
{\rm We thank the anonymous referee for a careful reading of the manuscript and for the useful comments and suggestions that greatly improved the paper}. Based on observations carried out under project number S19DL with the IRAM NOEMA Interferometer. IRAM is supported by INSU/CNRS (France), MPG (Germany) and IGN (Spain). We acknowledge IRAM staff for help provided during the observations and for data reduction. AP acknowledges support from Fondazione Cariplo grant no. 2020-0902. RAM acknowledges support from the ERC Advanced Grant 740246 (Cosmic Gas). This research made use of Astropy\footnote{http://www.astropy.org}, a community-developed core Python package for Astronomy \citep{AstropyI, AstropyII}, NumPy \citep{Numpy}, SciPy \citep{SciPy}, and Matplotlib \citep{Matplotlib}. 
\end{acknowledgements}
%
%
%
\bibliographystyle{aa}
\bibliography{MyBib}

\begin{appendix} 



\end{appendix}

\end{document}